\documentclass[aps,prb,twocolumn,groupedaddress]{revtex4-1}
\usepackage{amsmath}
\usepackage{graphicx}
\usepackage{color}

\newcommand{\ee}[0]{\mathrm{e}}
\newcommand{\kk}[0]{\mathbf{k}}
\newcommand{\qq}[0]{\mathbf{q}}
\newcommand{\rr}[0]{\mathbf{r}}

\begin{document}
\title{Collinear scattering of photoexcited carriers in graphene}
\author{Maxim Trushin}
\email{maxim.trushin@uni-konstanz.de}
\affiliation{Department of Physics, University of Konstanz, D-78457 Konstanz, Germany}

\begin{abstract}
We propose an explicitly solvable model for collinear scattering
of photoexcited carriers in intrinsic graphene irradiated by monochromatic light.
We find that the collinear scattering rate is directly proportional to the photocarrier energy and derive an analytic expression for the corresponding relaxation time.
The result agrees with the recent numerical prediction [Mihnev {\it et al.} Nat. Commun. {\bf 7}, 11617 (2016)]
and is able to describe the photocarrier evolution at low energies,
where scattering on optical phonons is strongly suppressed.
\end{abstract}

\pacs{}

\maketitle

\section{Introduction}

Graphene represents a single layer of carbon atoms exfoliated from bulk graphite \cite{PNAS2005novoselov} or grown by the chemical vapor deposition (CVD) technique.\cite{Natnano2010bae}
This material offers many extraordinary properties to exploit in optoelectronics such as universal (frequency-independent) absorption of light and overall low opacity,\cite{Science2008nair}
excellent electrical\cite{SSC2008bolotin} and thermal\cite{NanoLett2008balandin} conductivity.
Possible applications include transparent electrodes in displays and photovoltaic modules,\cite{ACS2010dearco}
high-speed electronic\cite{Science2010lin} and optical\cite{Nature2011liu} devices, energy storage\cite{Science2011zhu}, and many more.\cite{Nanoscale2015roadmap}
The unconventional optoelectronic properties of graphene are related to its extremely peculiar electronic spectrum:
The charge carriers demonstrate linear energy dispersion and there is no band gap between conduction and valence bands.\cite{McCann2012}
The combination of both is an exception among conducting materials\cite{Nature2007geim} and may offer innovative applications not possible within conventional approaches.
In particular, zero band gap means that the photocarriers can easily be excited even by using THz radiation,\cite{NanoLett2014jensen,PRB2013bacsi,PRL2016otto}
when optical phonon emission is strongly suppressed and
noncollinear carrier-carrier scattering turns out to be remarkably slow.\cite{PRL2016otto}
Our hypothesis is that the photocarrier evolution is governed by collinear
electron-electron (e-e) scattering, when all photocarrier momenta involved are {\em parallel}.\cite{PRB2007rana,PRBwinzer2012,PRB2013tomadin}
This is not possible in conventional semiconductors 
with the parabolic dispersion for carriers because momentum- and energy-balance equations cannot be satisfied in such processes simultaneously.
In contrast with previous approaches devoted to this problem,\cite{PRB2007rana,PRB2011kim,PRBwinzer2012,JAP2012pirro,Natcomm2013brida,PRB2013tomadin,PRB2013song,Natcomm2015mics,Natcomm2016mihnev}
the present work aims for an explicitly solvable model.

\begin{figure}
\includegraphics[width=\columnwidth]{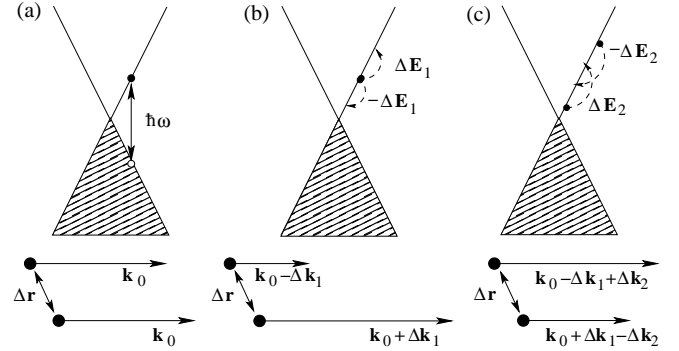}
\caption{\label{fig1} Collinear scattering in momentum and real space.
(a) Collinear photocarriers excited by monochromatic light with frequency $\omega$
have the same momentum $k_0$.
(b,c) Collinear photoelectrons  remain collinear while interacting with each other.
Most important, the distance between these electrons does not change,
because they travel with the same speed due to the linear dispersion.
Hence, the two-particle interaction potential $V(\rr)$ contributes at $\rr = \Delta\rr$ only
and can be taken equal to zero otherwise.}
\end{figure}

The reason why a collinear e-e scattering channel may dominate in photocarrier
relaxation is threefold.
First, pseudospin conservation results in partial suppression of e-e scattering with a non-zero scattering angle 
and even leads to complete suppression in the case of backscattering.\cite{PRB2013tomadin} This has been confirmed theoretically\cite{APL2012malic} and proven experimentally.\cite{Nanolett2014mittendorff,PRB2015trushin,PRL2016otto}
Second, collinear electrons remain collinear even though they may exchange momenta. Hence, they maintain a stable collinear scattering channel until a noncollinear electron 
comes into play. Third, thanks to the constant velocity, 
the distance between collinear electrons does not change while
they interact with each other. As a consequence, collinear carriers
``spend a lot of time together''\cite{PRB2013tomadin} and, therefore,
e-e interactions along graphene's conical bands are much more efficient than across.
Last but not least, there is a recent experimental evidence \cite{PRL2016otto}
of remarkably slow noncollinear e-e scattering 
obtained by means of pump-probe spectroscopy performed
at the excitation energies below the optical phonon energy.
These observations all together suggest that the full two-dimensional (2D) collision integral can be reduced to its one-dimensional (1D) analog
for the sake of simplicity.
In what follows, we derive an explicit formula for collinear relaxation time $\tau_\mathrm{coll}$ describing thermalization of photoelectrons excited 
by monochromatic light of frequency $\omega$, as shown in Fig.~\ref{fig1}.
The photocarrier relaxation rate reads
\begin{equation}
 \label{main-result}
 \frac{1}{\tau_\mathrm{coll}} = \frac{\tilde\alpha^2 E}{4\pi^3 \hbar}
\ln\left(\frac{1}{\tilde\alpha}\right),
\end{equation}
where $E$ is the photocarrier energy counted from the neutrality point, $\tilde\alpha=e^2/(\varepsilon \hbar v)$ is the effective fine-structure
constant for carriers in graphene determined by the electron charge $e$,
Plank constant $\hbar$, effective dielectric constant $\varepsilon$,
and carrier velocity $v$.
To derive Eq.~(\ref{main-result}) we employed
the renormalization procedure\cite{PRB2008fritz} based on the 
perturbation theory valid for $\tilde\alpha < 1$.
For graphene on the most conventional substrates, $\tilde\alpha$ is typically 
between $0.3$ and $0.8$; see Fig.~\ref{fig3}.
Hence, the photocarrier relaxation time scales with the photocarrier energy as 
$\tau_\mathrm{coll} \sim 1\,\mathrm{ps\cdot eV}/E$,
which is in perfect agreement with the recent
numerical result; see Fig. 6 in Ref. \onlinecite{Natcomm2016mihnev}. 
In Section \ref{model} we derive Eq.~(\ref{main-result}) and wrap up
in Section \ref{discussion} by discussing its physical consequences.


\section{Model}
\label{model}

\subsection{Preliminaries}

The carriers in graphene near the K-point of the first Brillouin zone are described by the massless Dirac Hamiltonian
$H_0=\hbar v \hat{\sigma}\cdot \mathbf{k}$, where $\hbar\mathbf{k}$ is the two-component
momentum operator, $\hat{\sigma}$ is the pseudospin operator constructed out of the Pauli matrices, and $v$ is the carrier velocity
determined by the tight-binding parameters for electrons on the honeycomb lattice.
The eigenstates of $H_0$ are given by $\varphi_{\kk s}(\rr)=\frac{1}{L\sqrt{2}}\ee^{i\kk\rr}(1,s\ee^{i\theta})^T$,
where $s=\pm 1$ is the band (or pseudospin) index, $\tan\theta = k_y/k_x$ is the direction of motion, $\rr$ is the two-component particle position, and  $L$ is the sample size.
The eigenvalues of $H_0$ are $E_s=s\hbar v k$.
The two-particle wave function can be constructed out of $\varphi_{\kk s}(\rr)$ as
\begin{equation}
 \psi_{\kk_i s_i \kk_j s_j}=\frac{1}{\sqrt{2}}\left[\varphi_{\kk_i s_i}(\rr_i)\varphi_{\kk_j s_j}(\rr_j) - \varphi_{\kk_i s_i}(\rr_j)\varphi_{\kk_j s_j}(\rr_i)\right].
\end{equation}
The probability for a given particle to occupy a given one-particle state with ($\kk,s$) is described by the distribution function $f_{\kk s}$ which satisfies the following 
differential equation\cite{PRB2013tomadin}
\begin{eqnarray}
\nonumber  \frac{d f_{\kk_1 s_1}}{dt} &= & \frac{2\pi}{\hbar}\sum\limits_{\kk_2,\kk_3,\kk_4} \sum\limits_{s_2,s_3,s_4} w(\kk_1 s_1,\kk_2 s_2;\kk_3 s_3,\kk_4 s_4) \\
\nonumber && \times \delta(E_{s_1}+E_{s_2}-E_{s_3}-E_{s_4}) \\
\nonumber && \times \left[(1-f_{\kk_1 s_1})(1-f_{\kk_2 s_2})f_{\kk_3 s_3} f_{\kk_4 s_4} \right. \\
 && \left. - f_{\kk_1 s_1} f_{\kk_2 s_2} (1-f_{\kk_3 s_3}) (1-f_{\kk_4 s_4})\right],
 \label{main-2d}
\end{eqnarray}
where $w(\kk_1 s_1,\kk_2 s_2;\kk_3 s_3,\kk_4 s_4)$ describes interaction of two particles and includes a direct (Hartree) and an exchange (Fock) term in the form \cite{PRB2013tomadin}
\begin{equation}
 w(\kk_1 s_1,\kk_2 s_2;\kk_3 s_3,\kk_4 s_4)=\frac{1}{2}\left| V_{1234} - V_{1243}\right|^2 + \left| V_{1234} \right|^2.
\end{equation}
Here, 
\begin{eqnarray}
\nonumber && V_{ijmn}= \frac{1}{4}\left(1+ s_i s_m \ee^{i(\theta_m-\theta_i)} \right) \left(1+ s_j s_n \ee^{i(\theta_n-\theta_j)} \right)  \\
&& \times \int \frac{d^2\rr_1}{L^2} \int \frac{d^2 \rr_2}{L^2} V(\rr_2-\rr_1) \ee^{i(\kk_m - \kk_i)\rr_1 + i(\kk_n - \kk_j)\rr_2}  
\label{Vijmn}
\end{eqnarray}
with $V(\rr)$ being the interaction potential.
The laws of pseudospin and momentum conservation are encoded in the first and second lines of Eq.~(\ref{Vijmn}) respectively.
Eq.~(\ref{main-2d}) is valid for an arbitrary $V(\rr)$ and contains information about collinear as well as noncollinear scattering.
It does not allow for an explicit solution in this general form, but provides the starting point for our model.

\subsection{Initial photocarrier distribution}

To solve Eq.~(\ref{main-2d}) we need an initial condition for $f_{\kk s}$. 
We suppose that $f_{\kk s}$ at $t=0$ is created by linearly polarized light and 
therefore anisotropic in momentum space.
The anisotropy has been predicted theoretically\cite{PRB2011malic,APL2012malic,EPL2011trushin,PRL2011trushin} 
and demonstrated experimentally.\cite{PRB2015trushin,Nanolett2014mittendorff,Nanolett2014echtermeyer}
The light-carrier interaction is described by the Hamiltonian $H_\mathrm{int}= \frac{e v}{c}\hat{\sigma}\cdot \mathbf{A}$,
where  $\mathbf{A}=\mathbf{A}_0\cos(\omega t- q z)$ is the vector potential created by the linearly polarized electromagnetic wave 
${\mathbf E}=\mathbf{E}_0\sin(\omega t- q z)$ with $\omega$ being the radiation frequency,
and $\mathbf{E}_0=\frac{\omega \mathbf{A}_0}{c}$. We assume normal incidence $\mathbf{q}\perp \mathbf{k}$ so that
there is no momentum transfer from photons to electrons. 
Due to the pseudospin selection rules\cite{PRB2015trushin}
the photocarrier momenta are aligned perpendicular to the polarization plane of light.
The e-e scattering also obeys pseudospin conservation and hence maintains anisotropy.\cite{APL2012malic}

It has been shown in Ref.~\onlinecite{PRB2015trushin} that the initial photoelectron distribution $f_{\kk s}(t=0)$ created by a monochromatic pump pulse 
in intrinsic graphene is given by $f_{\kk s}(0)=f_{\kk s}^{(0)}(0) + f_{\kk s}^{(1)}(0)$, where  
\begin{equation}
\label{eq0}
f_{\kk s}^{(0)}(0) = \frac{1}{1+\exp(s\hbar v k/T_0)}
\end{equation}
 is the Fermi-Dirac function at the initial temperature $T_0$ and zero chemical potential, and
\begin{eqnarray}
\label{sol3}
\nonumber f_{\kk s}^{(1)}(0) & = &
\frac{4\pi^2 \alpha v^2 \Phi}{\hbar \omega^2 }\sin^2(\theta-\theta_{E_0}) \delta(\omega - \Omega)\\
&& \times \left(f_{\kk(-s)}^{(0)}-f_{\kk(+s)}^{(0)}\right)
\end{eqnarray}
is the nonequilibrium addition. Here, $\Phi=(cE_0^2\delta t)/(8\pi)$ is the pump fluence with $\delta t$ being the pulse duration,
$\alpha=e^2/(\hbar c)$ is the fine-structure constant, $\hbar \Omega=2\hbar v k$ is the interband transition energy, and $\tan\theta_{E_0} = E_{0y}/E_{0x}$.
Since the initial temperature is low as compared with $\hbar\omega$ we can set $(f_{\kk(- s)}^{(0)}-f_{\kk(+s)}^{(0)}) = s$.
To take into account the finite spectral width of the pump pulse, the delta function can be substituted by a Gaussian distribution of the width $\Delta \omega$, i.e.,
$\delta(\omega - \Omega) \to \delta_{\Delta\omega}(\omega - \Omega)$, where
\begin{equation}
\label{gauss1}
 \delta_{\Delta\omega}(\omega - \Omega) = \frac{1}{\sqrt{2\pi}\Delta\omega}\mathrm{e}^{-\frac{(\omega - \Omega)^2}{2(\Delta\omega)^2}}.
\end{equation}

\subsection{Collinear limit}

In this subsection we simplify our model in order to 
investigate possible manifestations of collinear scattering.
We set a certain direction of motion in Eq.~(\ref{main-2d}) and employ dimensionless 1D momenta $\xi=k/k_0$, where
$k_0=\omega/2v$ is the central wave vector of photoexcited electrons and holes. 
Note, that  $-\infty <\xi < \infty$, and the energy dispersion is then given by $E_\xi=\hbar\omega \xi/2$.
There is no band index anymore because the conduction- and valence-band states are distinguished by the sign of $\xi$.
The initial distribution given by the sum of Eqs. (\ref{eq0}) and (\ref{sol3}) in the dimensionless units is written as $f_\xi(0)=f_\xi^{(0)}(0) + f_\xi^{(1)}(0)$, where
\begin{equation}
\label{eq00}
f_\xi^{(0)}(0) = \frac{1}{1+\exp(\beta_0 \xi)}, \quad  \beta_0=\frac{\hbar v k_0}{T_0},
\end{equation}
\begin{equation}
\label{sol33}
f_\xi^{(1)}(0)=\eta \left[\delta(1-\xi) - \delta(1+\xi)\right), \quad  \eta=\frac{n_\mathrm{ph}}{n_0}\sin^2(\theta-\theta_{E_0}).
\end{equation}
Here, $n_0=k_0^2/\pi$, and $n_\mathrm{ph}=\pi\alpha\Phi/(\hbar\omega)$ is the 2D photocarrier concentration
with $\pi\alpha$ being the linear optical absorption of graphene with the valley and spin degeneracy taken into account.
The first term in Eq.~(\ref{sol33}) corresponds to population of the conduction-band states, whereas the second one
describes depopulation of the valence-band states.
We can take into account the spectral width $\Delta \omega$
in a way similar to Eq.~(\ref{gauss1}), i.e., the delta-functions in Eq.~(\ref{sol33}) can be substituted by
\begin{equation}
\label{gauss2}
 \delta_\sigma(1 \pm \xi) = \frac{1}{\sqrt{2\pi}\sigma}\ee^{-\frac{(1 \pm \xi)^2}{2\sigma^2}},
\end{equation}
where $\sigma=\Delta \omega/ \omega$.

Since the distance between interacting collinear electrons does not change
we approximate the collinear interaction potential by a point-like one $V(\rr)=u_0\delta(\rr-\Delta\rr)$, where
$u_0$ is a constant independent of spatial coordinates,
and $\Delta\rr$ is the e-e mean distance that cancels out at the end of the day.
The scattering probability then reads
\begin{eqnarray}
\label{w1234}
&& w(\kk_1 s_1,\kk_2 s_2;\kk_3 s_3,\kk_4 s_4)=
\delta(\kk_3+\kk_4-\kk_1-\kk_2)\\
&& \nonumber \times \frac{\pi^2u_0^2}{L^6}
\left[1+s_1 s_3 \cos\left(\theta_3 -\theta_1\right)\right]
\left[1+s_2 s_4 \cos\left(\theta_4 -\theta_2\right)\right],
\end{eqnarray}
where the delta function represents momentum conservation.
Transforming the collision integral (\ref{main-2d}) into a 1D form is not a trivial task because the laws of energy and momentum conservation
result in the delta-function squared, which is not a well-defined function.
To overcome the difficulties associated with this divergence we employ the renormalization procedure\cite{PRB2008fritz} outlined in Appendix \ref{app1}. 
Finally, we transform the sums over $k_i$ in Eq.~(\ref{main-2d}) to the integrals over $\xi_i$ and obtain the following equation describing collinear scattering:
\begin{eqnarray}
\nonumber \frac{d f_{\xi_1}}{d\tau} & = &\int\limits_{-\infty}^\infty d\xi_3  \int\limits_{-\infty}^\infty d\xi_4  \sqrt{\frac{\xi_4 \xi_3 (\xi_3+\xi_4-\xi_1)}{\xi_1}} \\
\nonumber &&
\times \left[(1-f_{\xi_1})(1-f_{\xi_3+\xi_4-\xi_1})f_{\xi_3} f_{\xi_4} \right. \\
&& \left. - f_{\xi_1} f_{\xi_3+\xi_4-\xi_1} (1-f_{\xi_3}) (1-f_{\xi_4})\right],
 \label{main-1d-renorm}
\end{eqnarray}
where $\tau=t/t_0$ is the dimensionless time with $t_0^{-1}$ given by
\begin{equation}
 t_0^{-1}=\frac{u_0^2 k_0^3}{2\pi^3 \hbar^2 v}\ln\left(\frac{1}{\tilde\alpha}\right). 
 \label{t0}
\end{equation}

\subsection{Evolution of the photocarrier occupation}

\begin{figure}
\includegraphics[width=\columnwidth]{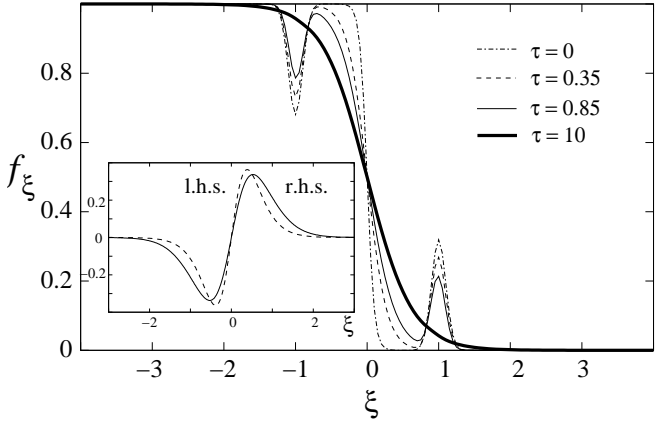}
\caption{\label{fig2} Evolution of the nonequilibrium distribution function $f_\xi(\tau)$ given by 
Eq.~(\ref{answer}) with $\eta=0.08$ corresponding to 
$\Phi=10\,\mathrm{\mu J/cm^2}$ at $\hbar\omega=1$ eV
or $\Phi=0.01\,\mathrm{\mu J/cm^2}$ at $\hbar\omega=100$ meV.
The delta-function is approximated by the Gaussian distribution (\ref{gauss2}) 
with $\sigma=0.1$, and the normal polarization configuration 
is assumed, i.e., $\theta-\theta_{E_0}=\pi/2$.
The carrier temperature rapidly increases with $\tau$ starting from $\beta_0=20$ 
(corresponds to the room temperature at $\hbar\omega=1$ eV)
to $\beta\sim 4$ ($T\sim 1450$ K). 
The inset shows that the right- and left-hand sides of Eq. (\ref{lhs-rhs-2}) are approximately equal at $\beta=4$.}
\end{figure}

Eq.~(\ref{main-1d-renorm}) is still too complicated for an analytic solution.
Since monochromatic radiation determines a characteristic photoelectron wave vector $k_0$
we assume that all momenta involved in collinear scattering are of the order of $k_0$, i.e., $\xi_i\sim 1$.
If we just set $\xi_i=1$ everywhere in the right-hand side of Eq.~(\ref{main-1d-renorm}), 
then the solution is trivially zero ($f_\xi=0$).
Therefore, we assume $\xi_i=1$ in the renormalization multiplier
so that the $\xi_i$-dependence is retained in the carrier occupation alone.
Hence, Eq.~(\ref{main-1d-renorm}) can be written as 
\begin{eqnarray}
\nonumber \frac{d f_{\xi_1}}{d\tau} & = &\int\limits_{-\infty}^\infty d\xi_3  \int\limits_{-\infty}^\infty d\xi_4 
\left[(1-f_{\xi_1})(1-f_{\xi_3+\xi_4-\xi_1})f_{\xi_3} f_{\xi_4} \right. \\
&& \left. - f_{\xi_1} f_{\xi_3+\xi_4-\xi_1} (1-f_{\xi_3}) (1-f_{\xi_4})\right].
 \label{main-1d}
\end{eqnarray}
This equation can be solved in the weak-excitation limit $\eta\ll 1$.
The {\it Ansatz} can be written as
$f_\xi(\tau)= f_\xi^{(0)}(\tau)+ f_\xi^{(1)}(\tau)$, where
\begin{equation}
 \label{probe20}
 f_\xi^{(0)}(\tau)=\frac{1}{1+\exp[\beta(\tau) \xi]},
\end{equation}
\begin{equation}
 \label{probe21}
f_\xi^{(1)}(\tau)= \eta \left[\delta(1-\xi) - \delta(1+\xi)\right]\ee^{-c(\xi)\tau}.
\end{equation}
Here, $\beta(\tau)$  and $c(\xi)$ are unknown functions to be determined.
Obviously $\beta(\tau)$ must satisfy the initial condition $\beta(0)=\beta_0$,
whereas $c(\xi)$ must be even $c(\xi)=c(-\xi)$.
Since the excitation is weak we neglect the terms of the order of $\eta^2$ and $\eta^3$, see Appendix \ref{app2}.
Assuming the low-temperature limit $\beta\gg 1$ we take the integrals over $\xi_{3,4}$ and write Eq.~(\ref{main-1d}) as
\begin{eqnarray}
\nonumber  && -\frac{\xi_1 (d\beta/d\tau) }{4\cosh^2(\beta\xi_1/2)} - \eta c(\xi_1)\left[\delta(1-\xi_1) - \delta(1+\xi_1)\right]\ee^{-c(\xi_1)\tau} \\
\nonumber &&  =  3\eta\left[\frac{-2}{1+\ee^{\beta \xi_1}} + \frac{\xi_1 -1}{\ee^{\beta(\xi_1-1)} -1} + \frac{\xi_1 +1}{1-\ee^{\beta(\xi_1+1)}}\right]\ee^{-c(1)\tau}\\
&& - \eta \frac{\xi_1^2}{2}\left[\delta(1-\xi_1) - \delta(1+\xi_1)\right]\ee^{-c(\xi_1)\tau},
\label{lhs-rhs}
\end{eqnarray}
Comparing the left- and right-hand sides in Eq.~(\ref{lhs-rhs}) we find that $c(\xi)=\xi^2/2$. In order figure out $\beta(\tau)$ we employ the energy-balance
equation for electrons which in our dimensionless units reads
\begin{equation}
 \label{balance}
 \int\limits_{0}^\infty d\xi \xi f_\xi (0) = \int\limits_{0}^\infty d\xi \xi f_\xi (\tau).
\end{equation}
The solution of Eq.~(\ref{balance}) is given by
\begin{equation}
\label{beta}
 \beta(\tau)= \frac{\beta_0}{\sqrt{1+ \frac{12\beta_0^2 \eta}{\pi^2}\left(1-\ee^{-\tau/2}\right)}}.
\end{equation}
The temperature for holes is the same as for electrons in intrinsic graphene.
Substituting Eq.~(\ref{beta}) and $c(\xi)=\xi^2/2$ into Eq.~(\ref{lhs-rhs}) the latter reduces to the following relation
\begin{equation}
\label{lhs-rhs-2}
 \frac{\xi_1 \beta^3}{4\pi^2\cosh^2(\beta\xi_1/2)} = \frac{-2}{1+\ee^{\beta \xi_1}} + \frac{\xi_1 -1}{\ee^{\beta(\xi_1-1)} -1} + \frac{\xi_1 +1}{1-\ee^{\beta(\xi_1+1)}},
\end{equation}
which approximately holds at reasonable $\beta>1$; see inset in Fig.~\ref{fig2}.
Indeed, both sides of Eq.~(\ref{lhs-rhs-2}) decay exponentially at $\beta\to\infty$ for $|\xi_1|>1$ and 
vanish completely at $\xi_1\to \pm \infty $ or $\xi_1=0$ for any finite $\beta$.
Hence, the approximate solution of Eq.~(\ref{main-1d}) at $\eta\ll 1$ and $\beta\gg 1$ reads
\begin{equation}
 \label{answer}
 f_{\xi}(\tau)= \frac{1}{1+\exp[\beta(\tau) \xi]} + \eta \left[\delta(1-\xi) - \delta(1+\xi)\right]\ee^{-\xi^2\frac{\tau}{2}},
\end{equation}
with $\beta(\tau)$ given by Eq.~(\ref{beta}). From Eq.~(\ref{answer}) one can deduce the dimensionless relaxation rate $\xi^2/2$
that approximately equals $1/2$ at $\xi\sim 1$.
The distribution function $f_{\xi}(\tau)$ is depicted in Fig. \ref{fig2} for
different $\tau$.

\section{Discussion and Conclusion}
\label{discussion}

\begin{figure}
\includegraphics[width=\columnwidth]{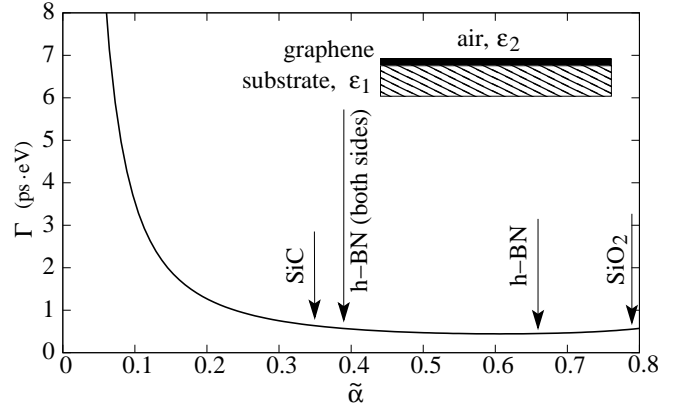}
\caption{\label{fig3} The photocarrier relaxation parameter (\ref{Gamma}) as 
a function of the effective fine-structure constant 
$\tilde\alpha=e^2/(\varepsilon \hbar v)$, where 
$\varepsilon=(\varepsilon_1+\varepsilon_2)/2$ is tunable by changing
the substrate. Arrows indicate positions for different substrates routinely
used in graphene optoelectronic devices.}
\end{figure}

The explicit solution (\ref{answer}) has been derived by assuming that $\xi\sim 1$, i.e., $k\sim k_0$.
Hence, all the quantities involved should also be evaluated having in mind this
approximation. In particular, $u_0$ should mimic the Fourier transform
of the Coulomb potential taken at $k\sim k_0$, i.e., $u_0=e^2/\varepsilon k_0$.
As consequence, $t_0$ takes the form
\begin{equation}
 t_0=\frac{4\pi^3}{\tilde\alpha^2 \omega \ln(1/\tilde\alpha)}. 
 \label{t0-final}
\end{equation}
In this approximation, the photocarrier energy is
$E=\hbar v k_0 \equiv \hbar\omega/2$, and the relaxation time deduced from 
Eq.~(\ref{answer}) reads $\tau_\mathrm{coll}=2 t_0$.
The result can be represented either by Eq.~(\ref{main-result}) or
in the form $\tau_\mathrm{coll}=\Gamma/E$, where $\Gamma$ is given by
\begin{equation}
\label{Gamma}
\Gamma=\frac{4\pi^3 \hbar}{\tilde\alpha^2 \ln(1/\tilde\alpha)},
\end{equation}
and shown in Fig.~\ref{fig3} for different
substrates. The effective dielectric constant $\varepsilon$
for graphene on a substrate is given by 
$\varepsilon=(\varepsilon_1+\varepsilon_2)/2$, where 
$\varepsilon_1$ and $\varepsilon_2$ are the relative permittivity
of the material below and above graphene layer, respectively.\cite{PRL2008jang}
For air, SiO$_2$, BN, and SiC the static relative permittivity 
is given by $1$, $3.9$, $5.06$, and $10.03$ respectively.\cite{PRL2008jang,PR1966geick,PRB1970lyle}
The carrier velocity is assumed to be $v=1.1\times 10^8\,\mathrm{cm/s}$.\cite{PRL2008jang}

The photocarrier relaxation time $\tau_\mathrm{coll}=\Gamma/E$ with
$\Gamma$ shown in Fig.~\ref{fig3} is the main result of this work.
Our outcomes are in perfect agreement with the recent numerical result,\cite{Natcomm2016mihnev} 
where calculations have been performed in full 2D momentum space
with all collinear and noncollinear e-e scattering processes included.
The numerical solution\cite{Natcomm2016mihnev}
suggests $\Gamma\approx 0.9\, \mathrm{eV\cdot ps}$, cf. Fig.~\ref{fig3} above.
Our model provides an {\it explicit} expression for this parameter.
Most important, the agreement between the full 2D model and our
approximate 1D approach suggests that collinear e-e collisions
{\it indeed} dominate photocarrier thermalization as long as the optical phonon emission is suppressed.
The most likely reason is the stability of collinear scattering channel explained in Introduction.

Our model can also qualitatively predict the evolution of a Gaussian photocarrier
distribution created by an impulsive optical excitation.
Since Eq.~(\ref{main-result}) suggests that the photocarriers with higher energies
thermalize faster, any initially symmetric Gaussian distribution
becomes asymmetric with the maximum shifted towards the neutrality point.
This is what one can see in Refs.~\onlinecite{PRB2013tomadin,APL2012malic}:
The electron distribution loses its Gaussian form and drifts 
towards the neutrality point while evolving in momentum space.
The particular reshaping of the photocarrier distribution
depends on the model for e-e screening.

The model proposed above can be assessed by measuring photocarrier evolution
using optical pump-probe spectroscopy well-established for graphene.\cite{APL2008dawlaty,PRL2010lui,PRB2011breusing,PRB2011hale,ACSnano2011shang,Natcomm2013brida,PRB2014yan,Nanolett2014mittendorff,PRB2015trushin,PRL2015gierz,PRL2016otto}
The most relevant experimental setup has been realized very recently
by K\"onig-Otto {\it et al.} in Ref.~\onlinecite{PRL2016otto},
where measurements have been performed at the excitation
energy below the optical phonon emission threshold.
Besides thermalization time, the hot carrier temperature can also be measured.
The temperature can be deduced from Eq.~(\ref{beta}) and in energy units reads
\begin{equation}
 \label{T}
 T_\mathrm{1D}=\sqrt{T_0^2 + \frac{12\alpha \hbar^2 v^2 \Phi}{\hbar\omega}\sin^2(\theta-\theta_{E_0}) \left(1-\ee^{-\frac{t}{2t_0}}\right)}.
\end{equation}
In the case of unpolarized light, $\sin^2(\theta-\theta_{E_0})$
should be substituted by $1/2$.
Eq.~(\ref{T}) in the limit $t\to\infty$ differs from the conventional estimation\cite{PRL2009bistritzer}
based on the energy-balance equation in intrinsic graphene neglecting phonon emission. 
In our notations, the conventional estimation can be written as
\begin{equation}
\label{T2d}
T_\mathrm{2D}=\left(T_0^3+\frac{\pi^2 \alpha \hbar^2 v^2 \Phi}{6\zeta(3)}\right)^{\frac{1}{3}}, 
\end{equation}
where $\zeta$ is the Riemann zeta function. The most striking difference between Eqs. (\ref{T}) and (\ref{T2d}) 
appears in the $\omega$-dependence: Eq. (\ref{T}) diverges at $\omega\to 0$, whereas Eq. (\ref{T2d}) does not.
The difference arises because, in our case, the photocarriers
are not yet thermalized over the whole two-dimensional momentum space.
For a more precise evaluation of the photocarrier temperature,
noncollinear e-e scattering and phonon emission 
should be taken into account.\cite{PRB2015trushin}

To conclude, we have developed an explicitly solvable model for photocarrier thermalization
due to collinear e-e scattering in graphene.
The model predicts $1/E$ scaling of collinear e-e relaxation time
that suggests its importance for thermalization in the high-energy tail of
a photoelectron distribution.
Note that this high-energy tail plays a leading role in thermionic emission
across silicon-graphene Schottky barriers\cite{NanoLett2016goykhman}
or graphene-isolator-graphene heterostructures.\cite{NatPhys2016ma,NanoLett2015rodriguez}
This paves the way towards correct assessment of the photocarrier thermalization dynamics in graphene-based optoelectronic devices.\cite{Bonaccorso2010review,Natnano2014review}

\acknowledgments

The author thanks Daniele Brida for fruitful discussions.

\appendix

\section{Collision integral in collinear limit}
\label{app1}

Here, we follow Ref.~\cite{PRB2008fritz} to write our collision integral in the collinear limit.
We start from Eqs. (\ref{main-2d}) and (\ref{w1234}), change the sums
to integrals as $\sum_{k_i}\to\int d^2k_i L^2/(2\pi)^2$, and make use of the
delta-function in Eq.~(\ref{w1234}) to integrate over $\kk_2$. The result reads
\begin{eqnarray}
\label{collin2}
\nonumber && \frac{d f_{k_1}}{d t} = \frac{u_0^2}{(2\pi)^3 \hbar^2 v}
\int dk_3 \int dk_3^\perp \int dk_4 \int dk_4^\perp\\
\nonumber && 
\times \delta\left(|\kk_1| + |\kk_3 + \kk_4 - \kk_1| - |\kk_3| - |\kk_4|\right) \\
\nonumber  &&
\times \left[(1-f_{k_1})(1-f_{k_3+k_4 -k_1})f_{k_3} f_{k_4} \right. \\
&& \left. 
- f_{k_1} f_{k_3+k_4 -k_1} (1-f_{k_3}) (1-f_{k_4})\right],
\end{eqnarray}
where $k_{3,4}^\perp$ are the components of $\kk_{3,4}$ perpendicular
to the collinear channel. They are set to zero in the distribution functions
as well as in the pseudospin form-factor.
It is convenient to introduce the variable $\qq=\kk_3 - \kk_1$
and assume that $\kk_1=(k_1,0)$, $\kk_4=(k_4,k_\perp)$, $\qq=(q,q_\perp)$.
Since the perpendicular components are small we can utilize the following approximate
expressions:
\begin{eqnarray}
&& |\kk_1+\qq|\approx k_1 + q + \frac{1}{2}\frac{q_\perp^2}{k_1+q},\\
&& |\kk_4+\qq|\approx k_4 + q + 
\frac{1}{2}\frac{\left(k_\perp + q_\perp \right)^2}{k_4 + q}, \\
&& |\kk_4|\approx k_4 + \frac{1}{2}\frac{k_\perp^2}{k_4}.
\end{eqnarray}
The delta-function can then be written as
\begin{eqnarray}
&& \delta\left(|\kk_1| + |\qq + \kk_4| - |\kk_1+\qq| - |\kk_4|\right)=\\
\nonumber && 2\delta\left(q_\perp^2 \frac{k_1-k_4}{(k_1+q)(k_4+q)}
+q_\perp \frac{2k_\perp}{k_4+q} - \frac{k_\perp^2 q}{k_4 (k_4+q)}\right)\\
\nonumber && =\left|\frac{2(k_4+q)(k_1+q)}{(k_1-k_4)(q_\perp^{(2)} -q_\perp^{(1)})}\right|
\left[\delta\left(q_\perp -q_\perp^{(1)}\right) 
+ \delta\left(q_\perp -q_\perp^{(2)}\right)\right],
\end{eqnarray}
where $q_\perp^{(1)}$ and $q_\perp^{(2)}$ are the roots of the argument of the delta-function.
The integral over $q_\perp$ becomes trivial, and Eq.~(\ref{collin2}) can be written as
\begin{eqnarray}
\label{collin3}
\nonumber && \frac{d f_{k_1}}{d t} = \frac{2 u_0^2}{(2\pi)^3 \hbar^2 v}
\int \frac{dk_\perp}{|k_\perp|} \\
\nonumber && 
\times \int dk_3 \int dk_4 \sqrt{\frac{k_4 k_3 (k_4+k_3 - k_1)}{k_1}} \\
\nonumber  &&
\times \left[(1-f_{k_1})(1-f_{k_3+k_4 -k_1})f_{k_3} f_{k_4} \right. \\
&& \left. 
- f_{k_1} f_{k_3+k_4 -k_1} (1-f_{k_3}) (1-f_{k_4})\right].
\end{eqnarray}
The integral over $k_\perp$ diverges but this divergence 
is cut off by self-energy corrections. As discussed in Ref.~\onlinecite{PRB2008fritz},
the important range of the $k_\perp$ integral is
between $T/(\hbar v)$ and $\tilde \alpha T/(\hbar v)$, which results in
$\int dk_\perp/k_\perp \approx 2\ln(1/\tilde\alpha)$.
Finally, we substitute $k_i$ and $t$ by corresponding dimensionless quantities
$\xi_i$ and $\tau$ and arrive at Eq.~(\ref{main-1d-renorm}).

\begin{widetext}
\section{One-dimensional collision integral in detail}
\label{app2}
Assuming that $f_\xi^{(0)}(\tau)$ and $f_\xi^{(1)}(\tau)$ are given by Eqs. (\ref{probe20}) and (\ref{probe21}), respectively, the collision integral in Eq.~(\ref{main-1d})
can be written as
\begin{eqnarray}
 \label{a1} && \int\limits_{-\infty}^\infty d\xi_3  \int\limits_{-\infty}^\infty d\xi_4 
 \left[(1-f_{\xi_1})(1-f_{\xi_3+\xi_4-\xi_1})f_{\xi_3} f_{\xi_4} - f_{\xi_1} f_{\xi_3+\xi_4-\xi_1} (1-f_{\xi_3}) (1-f_{\xi_4})\right] = \\
  \nonumber && \int\limits_{-\infty}^\infty d\xi_3  \int\limits_{-\infty}^\infty d\xi_4 \left\{
  f_{\xi_3}^{(1)}\left[\frac{1}{(1+\ee^{-\beta \xi_1})(1+\ee^{-\beta(\xi_3+\xi_4-\xi_1)})(1+\ee^{\beta\xi_4})} +
 \frac{1}{(1+\ee^{\beta \xi_1})(1+\ee^{\beta(\xi_3+\xi_4-\xi_1)})(1+\ee^{-\beta\xi_4})}\right] \right. + \\
  \nonumber && + f_{\xi_4}^{(1)}\left[\frac{1}{(1+\ee^{-\beta \xi_1})(1+\ee^{-\beta(\xi_3+\xi_4-\xi_1)})(1+\ee^{\beta\xi_3})} +
 \frac{1}{(1+\ee^{\beta \xi_1})(1+\ee^{\beta(\xi_3+\xi_4-\xi_1)})(1+\ee^{-\beta\xi_3})}\right] - \\
 \nonumber && - f_{\xi_1}^{(1)}\left[\frac{1}{(1+\ee^{-\beta(\xi_3+\xi_4-\xi_1)})(1+\ee^{\beta\xi_3})(1+\ee^{\beta\xi_4})} +
 \frac{1}{(1+\ee^{\beta(\xi_3+\xi_4-\xi_1)})(1+\ee^{-\beta\xi_3})(1+\ee^{-\beta\xi_4})}\right] -\\ 
 \nonumber && - \left. f_{\xi_3+\xi_4-\xi_1}^{(1)}\left[\frac{1}{(1+\ee^{-\beta\xi_1})(1+\ee^{\beta\xi_3})(1+\ee^{\beta\xi_4})} +
 \frac{1}{(1+\ee^{\beta\xi_1)})(1+\ee^{-\beta\xi_3})(1+\ee^{-\beta\xi_4})}\right] + o^2(\eta)\right\}.
\end{eqnarray}
Some integrals in Eq.~(\ref{a1}) are calculated by utilizing the $\delta(1\pm \xi_i)$-functions in $f_{\xi_i}$, and Eq.~(\ref{main-1d}) takes the form
\begin{eqnarray}
  \label{a2} && \frac{d f_{\xi_1}}{d\tau} = 3\eta \ee^{-c(1)\tau} \int\limits_{-\infty}^\infty \frac{d\xi_4 \sinh\beta}{\cosh\left[(\xi_4-\xi_1)\beta\right] + \cosh\beta}
 \left[\frac{1}{(1+\ee^{-\beta\xi_1})(1+\ee^{\beta\xi_4})} - \frac{1}{(1+\ee^{\beta\xi_1})(1+\ee^{-\beta\xi_4})}\right]\\
 && - f_{\xi_1}^{(1)} \int\limits_{-\infty}^\infty d\xi_3  \int\limits_{-\infty}^\infty d\xi_4 
 \left[\frac{1}{(1+\ee^{-\beta(\xi_3+\xi_4-\xi_1)})(1+\ee^{\beta\xi_3})(1+\ee^{\beta\xi_4})} +
 \frac{1}{(1+\ee^{\beta(\xi_3+\xi_4-\xi_1)})(1+\ee^{-\beta\xi_3})(1+\ee^{-\beta\xi_4})}\right].
 \nonumber
\end{eqnarray}
The first integral in Eq.~(\ref{a2}) can be calculated explicitly as
\begin{eqnarray}
\nonumber \int\limits_{-\infty}^\infty \frac{d\xi_4 \sinh\beta}{\cosh\left[(\xi_4-\xi_1)\beta\right] + \cosh\beta}
 \left[\frac{1}{(1+\ee^{-\beta\xi_1})(1+\ee^{\beta\xi_4})} - \frac{1}{(1+\ee^{\beta\xi_1})(1+\ee^{-\beta\xi_4})}\right]=
 \frac{-2}{1+\ee^{\beta \xi_1}} + \frac{\xi_1 -1}{\ee^{\beta(\xi_1-1)} -1} + \frac{\xi_1 +1}{1-\ee^{\beta(\xi_1+1)}}.
\end{eqnarray}
The second (double) integral can also be calculated analytically for $\xi_1>0$ and $\xi_1<0$ and the outcome represents a combination 
of logarithmic and polylogarithmic functions with the arguments containing $1- \ee^{\pm \beta\xi_1}$.
If we assume that $\beta \gg 1$, then $1-\ee^{\beta \xi_1}\approx -\ee^{\beta \xi_1}$ and $1-\ee^{-\beta \xi_1}\approx 1$ for $\xi_1>0$
and vice versa for $\xi_1<0$. In this approximation the integral takes a simple form
\begin{eqnarray}
\nonumber \int\limits_{-\infty}^\infty d\xi_3  \int\limits_{-\infty}^\infty d\xi_4 
 \left[\frac{1}{(1+\ee^{-\beta(\xi_3+\xi_4-\xi_1)})(1+\ee^{\beta\xi_3})(1+\ee^{\beta\xi_4})} +
 \frac{1}{(1+\ee^{\beta(\xi_3+\xi_4-\xi_1)})(1+\ee^{-\beta\xi_3})(1+\ee^{-\beta\xi_4})}\right]
 =\frac{\xi_1^2}{2}.
\end{eqnarray}
By using these expressions and calculating $\frac{d f_{\xi_1}}{d\tau}$ we transform Eq.~(\ref{a2}) to Eq.~(\ref{lhs-rhs}) in the main text.
\end{widetext}

\bibliography{collinear.bib,optical_pump_probe.bib,collinear-new.bib}

\begin{thebibliography}{48}%
\makeatletter
\providecommand \@ifxundefined [1]{%
 \@ifx{#1\undefined}
}%
\providecommand \@ifnum [1]{%
 \ifnum #1\expandafter \@firstoftwo
 \else \expandafter \@secondoftwo
 \fi
}%
\providecommand \@ifx [1]{%
 \ifx #1\expandafter \@firstoftwo
 \else \expandafter \@secondoftwo
 \fi
}%
\providecommand \natexlab [1]{#1}%
\providecommand \enquote  [1]{``#1''}%
\providecommand \bibnamefont  [1]{#1}%
\providecommand \bibfnamefont [1]{#1}%
\providecommand \citenamefont [1]{#1}%
\providecommand \href@noop [0]{\@secondoftwo}%
\providecommand \href [0]{\begingroup \@sanitize@url \@href}%
\providecommand \@href[1]{\@@startlink{#1}\@@href}%
\providecommand \@@href[1]{\endgroup#1\@@endlink}%
\providecommand \@sanitize@url [0]{\catcode `\\12\catcode `\$12\catcode
  `\&12\catcode `\#12\catcode `\^12\catcode `\_12\catcode `\%12\relax}%
\providecommand \@@startlink[1]{}%
\providecommand \@@endlink[0]{}%
\providecommand \url  [0]{\begingroup\@sanitize@url \@url }%
\providecommand \@url [1]{\endgroup\@href {#1}{\urlprefix }}%
\providecommand \urlprefix  [0]{URL }%
\providecommand \Eprint [0]{\href }%
\providecommand \doibase [0]{http://dx.doi.org/}%
\providecommand \selectlanguage [0]{\@gobble}%
\providecommand \bibinfo  [0]{\@secondoftwo}%
\providecommand \bibfield  [0]{\@secondoftwo}%
\providecommand \translation [1]{[#1]}%
\providecommand \BibitemOpen [0]{}%
\providecommand \bibitemStop [0]{}%
\providecommand \bibitemNoStop [0]{.\EOS\space}%
\providecommand \EOS [0]{\spacefactor3000\relax}%
\providecommand \BibitemShut  [1]{\csname bibitem#1\endcsname}%
\let\auto@bib@innerbib\@empty
\bibitem [{\citenamefont {Novoselov}\ \emph {et~al.}(2005)\citenamefont
  {Novoselov}, \citenamefont {Jiang}, \citenamefont {Schedin}, \citenamefont
  {Booth}, \citenamefont {Khotkevich}, \citenamefont {Morozov},\ and\
  \citenamefont {Geim}}]{PNAS2005novoselov}%
  \BibitemOpen
  \bibfield  {author} {\bibinfo {author} {\bibfnamefont {K.~S.}\ \bibnamefont
  {Novoselov}}, \bibinfo {author} {\bibfnamefont {D.}~\bibnamefont {Jiang}},
  \bibinfo {author} {\bibfnamefont {F.}~\bibnamefont {Schedin}}, \bibinfo
  {author} {\bibfnamefont {T.~J.}\ \bibnamefont {Booth}}, \bibinfo {author}
  {\bibfnamefont {V.~V.}\ \bibnamefont {Khotkevich}}, \bibinfo {author}
  {\bibfnamefont {S.~V.}\ \bibnamefont {Morozov}}, \ and\ \bibinfo {author}
  {\bibfnamefont {A.~K.}\ \bibnamefont {Geim}},\ }\href@noop {} {\bibfield
  {journal} {\bibinfo  {journal} {Proceedings of the National Academy of
  Sciences of the United States of America}\ }\textbf {\bibinfo {volume}
  {102}},\ \bibinfo {pages} {10451} (\bibinfo {year} {2005})}\BibitemShut
  {NoStop}%
\bibitem [{\citenamefont {Bae}\ \emph {et~al.}(2010)\citenamefont {Bae},
  \citenamefont {Kim}, \citenamefont {Lee}, \citenamefont {Xu}, \citenamefont
  {Park}, \citenamefont {Zheng}, \citenamefont {Balakrishnan}, \citenamefont
  {Lei}, \citenamefont {Ri~Kim}, \citenamefont {Song}, \citenamefont {Kim},
  \citenamefont {Kim}, \citenamefont {Ozyilmaz}, \citenamefont {Ahn},
  \citenamefont {Hong},\ and\ \citenamefont {Iijima}}]{Natnano2010bae}%
  \BibitemOpen
  \bibfield  {author} {\bibinfo {author} {\bibfnamefont {S.}~\bibnamefont
  {Bae}}, \bibinfo {author} {\bibfnamefont {H.}~\bibnamefont {Kim}}, \bibinfo
  {author} {\bibfnamefont {Y.}~\bibnamefont {Lee}}, \bibinfo {author}
  {\bibfnamefont {X.}~\bibnamefont {Xu}}, \bibinfo {author} {\bibfnamefont
  {J.-S.}\ \bibnamefont {Park}}, \bibinfo {author} {\bibfnamefont
  {Y.}~\bibnamefont {Zheng}}, \bibinfo {author} {\bibfnamefont
  {J.}~\bibnamefont {Balakrishnan}}, \bibinfo {author} {\bibfnamefont
  {T.}~\bibnamefont {Lei}}, \bibinfo {author} {\bibfnamefont {H.}~\bibnamefont
  {Ri~Kim}}, \bibinfo {author} {\bibfnamefont {Y.~I.}\ \bibnamefont {Song}},
  \bibinfo {author} {\bibfnamefont {Y.-J.}\ \bibnamefont {Kim}}, \bibinfo
  {author} {\bibfnamefont {K.~S.}\ \bibnamefont {Kim}}, \bibinfo {author}
  {\bibfnamefont {B.}~\bibnamefont {Ozyilmaz}}, \bibinfo {author}
  {\bibfnamefont {J.-H.}\ \bibnamefont {Ahn}}, \bibinfo {author} {\bibfnamefont
  {B.~H.}\ \bibnamefont {Hong}}, \ and\ \bibinfo {author} {\bibfnamefont
  {S.}~\bibnamefont {Iijima}},\ }\href@noop {} {\bibfield  {journal} {\bibinfo
  {journal} {Nat Nano}\ }\textbf {\bibinfo {volume} {5}},\ \bibinfo {pages}
  {574} (\bibinfo {year} {2010})}\BibitemShut {NoStop}%
\bibitem [{\citenamefont {Nair}\ \emph {et~al.}(2008)\citenamefont {Nair},
  \citenamefont {Blake}, \citenamefont {Grigorenko}, \citenamefont {Novoselov},
  \citenamefont {Booth}, \citenamefont {Stauber}, \citenamefont {Peres},\ and\
  \citenamefont {Geim}}]{Science2008nair}%
  \BibitemOpen
  \bibfield  {author} {\bibinfo {author} {\bibfnamefont {R.~R.}\ \bibnamefont
  {Nair}}, \bibinfo {author} {\bibfnamefont {P.}~\bibnamefont {Blake}},
  \bibinfo {author} {\bibfnamefont {A.~N.}\ \bibnamefont {Grigorenko}},
  \bibinfo {author} {\bibfnamefont {K.~S.}\ \bibnamefont {Novoselov}}, \bibinfo
  {author} {\bibfnamefont {T.~J.}\ \bibnamefont {Booth}}, \bibinfo {author}
  {\bibfnamefont {T.}~\bibnamefont {Stauber}}, \bibinfo {author} {\bibfnamefont
  {N.~M.~R.}\ \bibnamefont {Peres}}, \ and\ \bibinfo {author} {\bibfnamefont
  {A.~K.}\ \bibnamefont {Geim}},\ }\href@noop {} {\bibfield  {journal}
  {\bibinfo  {journal} {Science}\ }\textbf {\bibinfo {volume} {320}},\ \bibinfo
  {pages} {1308} (\bibinfo {year} {2008})}\BibitemShut {NoStop}%
\bibitem [{\citenamefont {Bolotin}\ \emph {et~al.}(2008)\citenamefont
  {Bolotin}, \citenamefont {Sikes}, \citenamefont {Jiang}, \citenamefont
  {Klima}, \citenamefont {Fudenberg}, \citenamefont {Hone}, \citenamefont
  {Kim},\ and\ \citenamefont {Stormer}}]{SSC2008bolotin}%
  \BibitemOpen
  \bibfield  {author} {\bibinfo {author} {\bibfnamefont {K.}~\bibnamefont
  {Bolotin}}, \bibinfo {author} {\bibfnamefont {K.}~\bibnamefont {Sikes}},
  \bibinfo {author} {\bibfnamefont {Z.}~\bibnamefont {Jiang}}, \bibinfo
  {author} {\bibfnamefont {M.}~\bibnamefont {Klima}}, \bibinfo {author}
  {\bibfnamefont {G.}~\bibnamefont {Fudenberg}}, \bibinfo {author}
  {\bibfnamefont {J.}~\bibnamefont {Hone}}, \bibinfo {author} {\bibfnamefont
  {P.}~\bibnamefont {Kim}}, \ and\ \bibinfo {author} {\bibfnamefont
  {H.}~\bibnamefont {Stormer}},\ }\href@noop {} {\bibfield  {journal} {\bibinfo
   {journal} {Solid State Communications}\ }\textbf {\bibinfo {volume} {146}},\
  \bibinfo {pages} {351 } (\bibinfo {year} {2008})}\BibitemShut {NoStop}%
\bibitem [{\citenamefont {Balandin}\ \emph {et~al.}(2008)\citenamefont
  {Balandin}, \citenamefont {Ghosh}, \citenamefont {Bao}, \citenamefont
  {Calizo}, \citenamefont {Teweldebrhan}, \citenamefont {Miao},\ and\
  \citenamefont {Lau}}]{NanoLett2008balandin}%
  \BibitemOpen
  \bibfield  {author} {\bibinfo {author} {\bibfnamefont {A.~A.}\ \bibnamefont
  {Balandin}}, \bibinfo {author} {\bibfnamefont {S.}~\bibnamefont {Ghosh}},
  \bibinfo {author} {\bibfnamefont {W.}~\bibnamefont {Bao}}, \bibinfo {author}
  {\bibfnamefont {I.}~\bibnamefont {Calizo}}, \bibinfo {author} {\bibfnamefont
  {D.}~\bibnamefont {Teweldebrhan}}, \bibinfo {author} {\bibfnamefont
  {F.}~\bibnamefont {Miao}}, \ and\ \bibinfo {author} {\bibfnamefont {C.~N.}\
  \bibnamefont {Lau}},\ }\href@noop {} {\bibfield  {journal} {\bibinfo
  {journal} {Nano Letters}\ }\textbf {\bibinfo {volume} {8}},\ \bibinfo {pages}
  {902} (\bibinfo {year} {2008})}\BibitemShut {NoStop}%
\bibitem [{\citenamefont {Arco}\ \emph {et~al.}(2010)\citenamefont {Arco},
  \citenamefont {Zhang}, \citenamefont {Schlenker}, \citenamefont {Ryu},
  \citenamefont {Thompson},\ and\ \citenamefont {Zhou}}]{ACS2010dearco}%
  \BibitemOpen
  \bibfield  {author} {\bibinfo {author} {\bibfnamefont {L.~G.~D.}\
  \bibnamefont {Arco}}, \bibinfo {author} {\bibfnamefont {Y.}~\bibnamefont
  {Zhang}}, \bibinfo {author} {\bibfnamefont {C.~W.}\ \bibnamefont
  {Schlenker}}, \bibinfo {author} {\bibfnamefont {K.}~\bibnamefont {Ryu}},
  \bibinfo {author} {\bibfnamefont {M.~E.}\ \bibnamefont {Thompson}}, \ and\
  \bibinfo {author} {\bibfnamefont {C.}~\bibnamefont {Zhou}},\ }\href@noop {}
  {\bibfield  {journal} {\bibinfo  {journal} {ACS Nano}\ }\textbf {\bibinfo
  {volume} {4}},\ \bibinfo {pages} {2865} (\bibinfo {year} {2010})}\BibitemShut
  {NoStop}%
\bibitem [{\citenamefont {Lin}\ \emph {et~al.}(2010)\citenamefont {Lin},
  \citenamefont {Dimitrakopoulos}, \citenamefont {Jenkins}, \citenamefont
  {Farmer}, \citenamefont {Chiu}, \citenamefont {Grill},\ and\ \citenamefont
  {Avouris}}]{Science2010lin}%
  \BibitemOpen
  \bibfield  {author} {\bibinfo {author} {\bibfnamefont {Y.-M.}\ \bibnamefont
  {Lin}}, \bibinfo {author} {\bibfnamefont {C.}~\bibnamefont
  {Dimitrakopoulos}}, \bibinfo {author} {\bibfnamefont {K.~A.}\ \bibnamefont
  {Jenkins}}, \bibinfo {author} {\bibfnamefont {D.~B.}\ \bibnamefont {Farmer}},
  \bibinfo {author} {\bibfnamefont {H.-Y.}\ \bibnamefont {Chiu}}, \bibinfo
  {author} {\bibfnamefont {A.}~\bibnamefont {Grill}}, \ and\ \bibinfo {author}
  {\bibfnamefont {P.}~\bibnamefont {Avouris}},\ }\href@noop {} {\bibfield
  {journal} {\bibinfo  {journal} {Science}\ }\textbf {\bibinfo {volume}
  {327}},\ \bibinfo {pages} {662} (\bibinfo {year} {2010})}\BibitemShut
  {NoStop}%
\bibitem [{\citenamefont {Liu}\ \emph {et~al.}(2011)\citenamefont {Liu},
  \citenamefont {Yin}, \citenamefont {Ulin-Avila}, \citenamefont {Geng},
  \citenamefont {Zentgraf}, \citenamefont {Ju}, \citenamefont {Wang},\ and\
  \citenamefont {Zhang}}]{Nature2011liu}%
  \BibitemOpen
  \bibfield  {author} {\bibinfo {author} {\bibfnamefont {M.}~\bibnamefont
  {Liu}}, \bibinfo {author} {\bibfnamefont {X.}~\bibnamefont {Yin}}, \bibinfo
  {author} {\bibfnamefont {E.}~\bibnamefont {Ulin-Avila}}, \bibinfo {author}
  {\bibfnamefont {B.}~\bibnamefont {Geng}}, \bibinfo {author} {\bibfnamefont
  {T.}~\bibnamefont {Zentgraf}}, \bibinfo {author} {\bibfnamefont
  {L.}~\bibnamefont {Ju}}, \bibinfo {author} {\bibfnamefont {F.}~\bibnamefont
  {Wang}}, \ and\ \bibinfo {author} {\bibfnamefont {X.}~\bibnamefont {Zhang}},\
  }\href@noop {} {\bibfield  {journal} {\bibinfo  {journal} {Nature}\ }\textbf
  {\bibinfo {volume} {474}},\ \bibinfo {pages} {64} (\bibinfo {year}
  {2011})}\BibitemShut {NoStop}%
\bibitem [{\citenamefont {Zhu}\ \emph {et~al.}(2011)\citenamefont {Zhu},
  \citenamefont {Murali}, \citenamefont {Stoller}, \citenamefont {Ganesh},
  \citenamefont {Cai}, \citenamefont {Ferreira}, \citenamefont {Pirkle},
  \citenamefont {Wallace}, \citenamefont {Cychosz}, \citenamefont {Thommes},
  \citenamefont {Su}, \citenamefont {Stach},\ and\ \citenamefont
  {Ruoff}}]{Science2011zhu}%
  \BibitemOpen
  \bibfield  {author} {\bibinfo {author} {\bibfnamefont {Y.}~\bibnamefont
  {Zhu}}, \bibinfo {author} {\bibfnamefont {S.}~\bibnamefont {Murali}},
  \bibinfo {author} {\bibfnamefont {M.~D.}\ \bibnamefont {Stoller}}, \bibinfo
  {author} {\bibfnamefont {K.~J.}\ \bibnamefont {Ganesh}}, \bibinfo {author}
  {\bibfnamefont {W.}~\bibnamefont {Cai}}, \bibinfo {author} {\bibfnamefont
  {P.~J.}\ \bibnamefont {Ferreira}}, \bibinfo {author} {\bibfnamefont
  {A.}~\bibnamefont {Pirkle}}, \bibinfo {author} {\bibfnamefont {R.~M.}\
  \bibnamefont {Wallace}}, \bibinfo {author} {\bibfnamefont {K.~A.}\
  \bibnamefont {Cychosz}}, \bibinfo {author} {\bibfnamefont {M.}~\bibnamefont
  {Thommes}}, \bibinfo {author} {\bibfnamefont {D.}~\bibnamefont {Su}},
  \bibinfo {author} {\bibfnamefont {E.~A.}\ \bibnamefont {Stach}}, \ and\
  \bibinfo {author} {\bibfnamefont {R.~S.}\ \bibnamefont {Ruoff}},\ }\href@noop
  {} {\bibfield  {journal} {\bibinfo  {journal} {Science}\ }\textbf {\bibinfo
  {volume} {332}},\ \bibinfo {pages} {1537} (\bibinfo {year}
  {2011})}\BibitemShut {NoStop}%
\bibitem [{\citenamefont {Ferrari}\ \emph {et~al.}(2015)\citenamefont
  {Ferrari}, \citenamefont {Bonaccorso}, \citenamefont {Fal{'}ko},
  \citenamefont {Novoselov}, \citenamefont {Roche}, \citenamefont {Boggild},
  \citenamefont {Borini}, \citenamefont {Koppens}, \citenamefont {Palermo},
  \citenamefont {Pugno}, \citenamefont {Garrido}, \citenamefont {Sordan},
  \citenamefont {Bianco}, \citenamefont {Ballerini}, \citenamefont {Prato},
  \citenamefont {Lidorikis}, \citenamefont {Kivioja}, \citenamefont
  {Marinelli}, \citenamefont {Ryhanen}, \citenamefont {Morpurgo}, \citenamefont
  {Coleman}, \citenamefont {Nicolosi}, \citenamefont {Colombo}, \citenamefont
  {Fert}, \citenamefont {Garcia-Hernandez}, \citenamefont {Bachtold},
  \citenamefont {Schneider}, \citenamefont {Guinea}, \citenamefont {Dekker},
  \citenamefont {Barbone}, \citenamefont {Sun}, \citenamefont {Galiotis},
  \citenamefont {Grigorenko}, \citenamefont {Konstantatos}, \citenamefont
  {Kis}, \citenamefont {Katsnelson}, \citenamefont {Vandersypen}, \citenamefont
  {Loiseau}, \citenamefont {Morandi}, \citenamefont {Neumaier}, \citenamefont
  {Treossi}, \citenamefont {Pellegrini}, \citenamefont {Polini}, \citenamefont
  {Tredicucci}, \citenamefont {Williams}, \citenamefont {Hee~Hong},
  \citenamefont {Ahn}, \citenamefont {Min~Kim}, \citenamefont {Zirath},
  \citenamefont {van Wees}, \citenamefont {van~der Zant}, \citenamefont
  {Occhipinti}, \citenamefont {Di~Matteo}, \citenamefont {Kinloch},
  \citenamefont {Seyller}, \citenamefont {Quesnel}, \citenamefont {Feng},
  \citenamefont {Teo}, \citenamefont {Rupesinghe}, \citenamefont {Hakonen},
  \citenamefont {Neil}, \citenamefont {Tannock}, \citenamefont {Lofwander},\
  and\ \citenamefont {Kinaret}}]{Nanoscale2015roadmap}%
  \BibitemOpen
  \bibfield  {author} {\bibinfo {author} {\bibfnamefont {A.~C.}\ \bibnamefont
  {Ferrari}}, \bibinfo {author} {\bibfnamefont {F.}~\bibnamefont {Bonaccorso}},
  \bibinfo {author} {\bibfnamefont {V.}~\bibnamefont {Fal{'}ko}}, \bibinfo
  {author} {\bibfnamefont {K.~S.}\ \bibnamefont {Novoselov}}, \bibinfo {author}
  {\bibfnamefont {S.}~\bibnamefont {Roche}}, \bibinfo {author} {\bibfnamefont
  {P.}~\bibnamefont {Boggild}}, \bibinfo {author} {\bibfnamefont
  {S.}~\bibnamefont {Borini}}, \bibinfo {author} {\bibfnamefont {F.~H.~L.}\
  \bibnamefont {Koppens}}, \bibinfo {author} {\bibfnamefont {V.}~\bibnamefont
  {Palermo}}, \bibinfo {author} {\bibfnamefont {N.}~\bibnamefont {Pugno}},
  \bibinfo {author} {\bibfnamefont {J.~A.}\ \bibnamefont {Garrido}}, \bibinfo
  {author} {\bibfnamefont {R.}~\bibnamefont {Sordan}}, \bibinfo {author}
  {\bibfnamefont {A.}~\bibnamefont {Bianco}}, \bibinfo {author} {\bibfnamefont
  {L.}~\bibnamefont {Ballerini}}, \bibinfo {author} {\bibfnamefont
  {M.}~\bibnamefont {Prato}}, \bibinfo {author} {\bibfnamefont
  {E.}~\bibnamefont {Lidorikis}}, \bibinfo {author} {\bibfnamefont
  {J.}~\bibnamefont {Kivioja}}, \bibinfo {author} {\bibfnamefont
  {C.}~\bibnamefont {Marinelli}}, \bibinfo {author} {\bibfnamefont
  {T.}~\bibnamefont {Ryhanen}}, \bibinfo {author} {\bibfnamefont
  {A.}~\bibnamefont {Morpurgo}}, \bibinfo {author} {\bibfnamefont {J.~N.}\
  \bibnamefont {Coleman}}, \bibinfo {author} {\bibfnamefont {V.}~\bibnamefont
  {Nicolosi}}, \bibinfo {author} {\bibfnamefont {L.}~\bibnamefont {Colombo}},
  \bibinfo {author} {\bibfnamefont {A.}~\bibnamefont {Fert}}, \bibinfo {author}
  {\bibfnamefont {M.}~\bibnamefont {Garcia-Hernandez}}, \bibinfo {author}
  {\bibfnamefont {A.}~\bibnamefont {Bachtold}}, \bibinfo {author}
  {\bibfnamefont {G.~F.}\ \bibnamefont {Schneider}}, \bibinfo {author}
  {\bibfnamefont {F.}~\bibnamefont {Guinea}}, \bibinfo {author} {\bibfnamefont
  {C.}~\bibnamefont {Dekker}}, \bibinfo {author} {\bibfnamefont
  {M.}~\bibnamefont {Barbone}}, \bibinfo {author} {\bibfnamefont
  {Z.}~\bibnamefont {Sun}}, \bibinfo {author} {\bibfnamefont {C.}~\bibnamefont
  {Galiotis}}, \bibinfo {author} {\bibfnamefont {A.~N.}\ \bibnamefont
  {Grigorenko}}, \bibinfo {author} {\bibfnamefont {G.}~\bibnamefont
  {Konstantatos}}, \bibinfo {author} {\bibfnamefont {A.}~\bibnamefont {Kis}},
  \bibinfo {author} {\bibfnamefont {M.}~\bibnamefont {Katsnelson}}, \bibinfo
  {author} {\bibfnamefont {L.}~\bibnamefont {Vandersypen}}, \bibinfo {author}
  {\bibfnamefont {A.}~\bibnamefont {Loiseau}}, \bibinfo {author} {\bibfnamefont
  {V.}~\bibnamefont {Morandi}}, \bibinfo {author} {\bibfnamefont
  {D.}~\bibnamefont {Neumaier}}, \bibinfo {author} {\bibfnamefont
  {E.}~\bibnamefont {Treossi}}, \bibinfo {author} {\bibfnamefont
  {V.}~\bibnamefont {Pellegrini}}, \bibinfo {author} {\bibfnamefont
  {M.}~\bibnamefont {Polini}}, \bibinfo {author} {\bibfnamefont
  {A.}~\bibnamefont {Tredicucci}}, \bibinfo {author} {\bibfnamefont {G.~M.}\
  \bibnamefont {Williams}}, \bibinfo {author} {\bibfnamefont {B.}~\bibnamefont
  {Hee~Hong}}, \bibinfo {author} {\bibfnamefont {J.-H.}\ \bibnamefont {Ahn}},
  \bibinfo {author} {\bibfnamefont {J.}~\bibnamefont {Min~Kim}}, \bibinfo
  {author} {\bibfnamefont {H.}~\bibnamefont {Zirath}}, \bibinfo {author}
  {\bibfnamefont {B.~J.}\ \bibnamefont {van Wees}}, \bibinfo {author}
  {\bibfnamefont {H.}~\bibnamefont {van~der Zant}}, \bibinfo {author}
  {\bibfnamefont {L.}~\bibnamefont {Occhipinti}}, \bibinfo {author}
  {\bibfnamefont {A.}~\bibnamefont {Di~Matteo}}, \bibinfo {author}
  {\bibfnamefont {I.~A.}\ \bibnamefont {Kinloch}}, \bibinfo {author}
  {\bibfnamefont {T.}~\bibnamefont {Seyller}}, \bibinfo {author} {\bibfnamefont
  {E.}~\bibnamefont {Quesnel}}, \bibinfo {author} {\bibfnamefont
  {X.}~\bibnamefont {Feng}}, \bibinfo {author} {\bibfnamefont {K.}~\bibnamefont
  {Teo}}, \bibinfo {author} {\bibfnamefont {N.}~\bibnamefont {Rupesinghe}},
  \bibinfo {author} {\bibfnamefont {P.}~\bibnamefont {Hakonen}}, \bibinfo
  {author} {\bibfnamefont {S.~R.~T.}\ \bibnamefont {Neil}}, \bibinfo {author}
  {\bibfnamefont {Q.}~\bibnamefont {Tannock}}, \bibinfo {author} {\bibfnamefont
  {T.}~\bibnamefont {Lofwander}}, \ and\ \bibinfo {author} {\bibfnamefont
  {J.}~\bibnamefont {Kinaret}},\ }\href@noop {} {\bibfield  {journal} {\bibinfo
   {journal} {Nanoscale}\ }\textbf {\bibinfo {volume} {7}},\ \bibinfo {pages}
  {4598} (\bibinfo {year} {2015})}\BibitemShut {NoStop}%
\bibitem [{\citenamefont {McCann­}(2012)}]{McCann2012}%
  \BibitemOpen
  \bibfield  {author} {\bibinfo {author} {\bibfnamefont {E.}~\bibnamefont
  {McCann­}},\ }\href@noop {} {\emph {\bibinfo {title} {Graphene
  Nanoelectronics: Metrology, Synthesis, Properties and Applications, chapter
  8}}}\ (\bibinfo  {publisher} {Springer-Verlag Berlin Heidelberg},\ \bibinfo
  {year} {2012})\ pp.\ \bibinfo {pages} {pages 237--275}\BibitemShut {NoStop}%
\bibitem [{\citenamefont {Geim}\ and\ \citenamefont
  {Novoselov}(2007)}]{Nature2007geim}%
  \BibitemOpen
  \bibfield  {author} {\bibinfo {author} {\bibfnamefont {A.~K.}\ \bibnamefont
  {Geim}}\ and\ \bibinfo {author} {\bibfnamefont {K.~S.}\ \bibnamefont
  {Novoselov}},\ }\href@noop {} {\bibfield  {journal} {\bibinfo  {journal}
  {Nat. Mat.}\ }\textbf {\bibinfo {volume} {6}},\ \bibinfo {pages} {183}
  (\bibinfo {year} {2007})}\BibitemShut {NoStop}%
\bibitem [{\citenamefont {Jensen}\ \emph {et~al.}(2014)\citenamefont {Jensen},
  \citenamefont {Mics}, \citenamefont {Ivanov}, \citenamefont {Varol},
  \citenamefont {Turchinovich}, \citenamefont {Koppens}, \citenamefont {Bonn},\
  and\ \citenamefont {Tielrooij}}]{NanoLett2014jensen}%
  \BibitemOpen
  \bibfield  {author} {\bibinfo {author} {\bibfnamefont {S.~A.}\ \bibnamefont
  {Jensen}}, \bibinfo {author} {\bibfnamefont {Z.}~\bibnamefont {Mics}},
  \bibinfo {author} {\bibfnamefont {I.}~\bibnamefont {Ivanov}}, \bibinfo
  {author} {\bibfnamefont {H.~S.}\ \bibnamefont {Varol}}, \bibinfo {author}
  {\bibfnamefont {D.}~\bibnamefont {Turchinovich}}, \bibinfo {author}
  {\bibfnamefont {F.~H.~L.}\ \bibnamefont {Koppens}}, \bibinfo {author}
  {\bibfnamefont {M.}~\bibnamefont {Bonn}}, \ and\ \bibinfo {author}
  {\bibfnamefont {K.~J.}\ \bibnamefont {Tielrooij}},\ }\href@noop {} {\bibfield
   {journal} {\bibinfo  {journal} {Nano Letters}\ }\textbf {\bibinfo {volume}
  {14}},\ \bibinfo {pages} {5839} (\bibinfo {year} {2014})}\BibitemShut
  {NoStop}%
\bibitem [{\citenamefont {Bacsi}\ and\ \citenamefont
  {Virosztek}(2013)}]{PRB2013bacsi}%
  \BibitemOpen
  \bibfield  {author} {\bibinfo {author} {\bibfnamefont {A.}~\bibnamefont
  {Bacsi}}\ and\ \bibinfo {author} {\bibfnamefont {A.}~\bibnamefont
  {Virosztek}},\ }\href@noop {} {\bibfield  {journal} {\bibinfo  {journal}
  {Phys. Rev. B}\ }\textbf {\bibinfo {volume} {87}},\ \bibinfo {pages} {125425}
  (\bibinfo {year} {2013})}\BibitemShut {NoStop}%
\bibitem [{\citenamefont {K\"onig-Otto}\ \emph {et~al.}(2016)\citenamefont
  {K\"onig-Otto}, \citenamefont {Mittendorff}, \citenamefont {Winzer},
  \citenamefont {Kadi}, \citenamefont {Malic}, \citenamefont {Knorr},
  \citenamefont {Berger}, \citenamefont {de~Heer}, \citenamefont {Pashkin},
  \citenamefont {Schneider}, \citenamefont {Helm},\ and\ \citenamefont
  {Winnerl}}]{PRL2016otto}%
  \BibitemOpen
  \bibfield  {author} {\bibinfo {author} {\bibfnamefont {J.~C.}\ \bibnamefont
  {K\"onig-Otto}}, \bibinfo {author} {\bibfnamefont {M.}~\bibnamefont
  {Mittendorff}}, \bibinfo {author} {\bibfnamefont {T.}~\bibnamefont {Winzer}},
  \bibinfo {author} {\bibfnamefont {F.}~\bibnamefont {Kadi}}, \bibinfo {author}
  {\bibfnamefont {E.}~\bibnamefont {Malic}}, \bibinfo {author} {\bibfnamefont
  {A.}~\bibnamefont {Knorr}}, \bibinfo {author} {\bibfnamefont
  {C.}~\bibnamefont {Berger}}, \bibinfo {author} {\bibfnamefont {W.~A.}\
  \bibnamefont {de~Heer}}, \bibinfo {author} {\bibfnamefont {A.}~\bibnamefont
  {Pashkin}}, \bibinfo {author} {\bibfnamefont {H.}~\bibnamefont {Schneider}},
  \bibinfo {author} {\bibfnamefont {M.}~\bibnamefont {Helm}}, \ and\ \bibinfo
  {author} {\bibfnamefont {S.}~\bibnamefont {Winnerl}},\ }\href@noop {}
  {\bibfield  {journal} {\bibinfo  {journal} {Phys. Rev. Lett.}\ }\textbf
  {\bibinfo {volume} {117}},\ \bibinfo {pages} {087401} (\bibinfo {year}
  {2016})}\BibitemShut {NoStop}%
\bibitem [{\citenamefont {Rana}(2007)}]{PRB2007rana}%
  \BibitemOpen
  \bibfield  {author} {\bibinfo {author} {\bibfnamefont {F.}~\bibnamefont
  {Rana}},\ }\href@noop {} {\bibfield  {journal} {\bibinfo  {journal} {Phys.
  Rev. B}\ }\textbf {\bibinfo {volume} {76}},\ \bibinfo {pages} {155431}
  (\bibinfo {year} {2007})}\BibitemShut {NoStop}%
\bibitem [{\citenamefont {Winzer}\ and\ \citenamefont
  {Mali\ifmmode~\acute{c}\else \'{c}\fi{}}(2012)}]{PRBwinzer2012}%
  \BibitemOpen
  \bibfield  {author} {\bibinfo {author} {\bibfnamefont {T.}~\bibnamefont
  {Winzer}}\ and\ \bibinfo {author} {\bibfnamefont {E.}~\bibnamefont
  {Mali\ifmmode~\acute{c}\else \'{c}\fi{}}},\ }\href@noop {} {\bibfield
  {journal} {\bibinfo  {journal} {Phys. Rev. B}\ }\textbf {\bibinfo {volume}
  {85}},\ \bibinfo {pages} {241404} (\bibinfo {year} {2012})}\BibitemShut
  {NoStop}%
\bibitem [{\citenamefont {Tomadin}\ \emph {et~al.}(2013)\citenamefont
  {Tomadin}, \citenamefont {Brida}, \citenamefont {Cerullo}, \citenamefont
  {Ferrari},\ and\ \citenamefont {Polini}}]{PRB2013tomadin}%
  \BibitemOpen
  \bibfield  {author} {\bibinfo {author} {\bibfnamefont {A.}~\bibnamefont
  {Tomadin}}, \bibinfo {author} {\bibfnamefont {D.}~\bibnamefont {Brida}},
  \bibinfo {author} {\bibfnamefont {G.}~\bibnamefont {Cerullo}}, \bibinfo
  {author} {\bibfnamefont {A.~C.}\ \bibnamefont {Ferrari}}, \ and\ \bibinfo
  {author} {\bibfnamefont {M.}~\bibnamefont {Polini}},\ }\href@noop {}
  {\bibfield  {journal} {\bibinfo  {journal} {Phys. Rev. B}\ }\textbf {\bibinfo
  {volume} {88}},\ \bibinfo {pages} {035430} (\bibinfo {year}
  {2013})}\BibitemShut {NoStop}%
\bibitem [{\citenamefont {Kim}\ \emph {et~al.}(2011)\citenamefont {Kim},
  \citenamefont {Perebeinos},\ and\ \citenamefont {Avouris}}]{PRB2011kim}%
  \BibitemOpen
  \bibfield  {author} {\bibinfo {author} {\bibfnamefont {R.}~\bibnamefont
  {Kim}}, \bibinfo {author} {\bibfnamefont {V.}~\bibnamefont {Perebeinos}}, \
  and\ \bibinfo {author} {\bibfnamefont {P.}~\bibnamefont {Avouris}},\
  }\href@noop {} {\bibfield  {journal} {\bibinfo  {journal} {Phys. Rev. B}\
  }\textbf {\bibinfo {volume} {84}},\ \bibinfo {pages} {075449} (\bibinfo
  {year} {2011})}\BibitemShut {NoStop}%
\bibitem [{\citenamefont {Pirro}\ \emph {et~al.}(2012)\citenamefont {Pirro},
  \citenamefont {Girdhar}, \citenamefont {Leblebici},\ and\ \citenamefont
  {Leburton}}]{JAP2012pirro}%
  \BibitemOpen
  \bibfield  {author} {\bibinfo {author} {\bibfnamefont {L.}~\bibnamefont
  {Pirro}}, \bibinfo {author} {\bibfnamefont {A.}~\bibnamefont {Girdhar}},
  \bibinfo {author} {\bibfnamefont {Y.}~\bibnamefont {Leblebici}}, \ and\
  \bibinfo {author} {\bibfnamefont {J.-P.}\ \bibnamefont {Leburton}},\
  }\href@noop {} {\bibfield  {journal} {\bibinfo  {journal} {Journal of Applied
  Physics}\ }\textbf {\bibinfo {volume} {112}},\ \bibinfo {eid} {093707}
  (\bibinfo {year} {2012})}\BibitemShut {NoStop}%
\bibitem [{\citenamefont {Brida}\ \emph {et~al.}(2013)\citenamefont {Brida},
  \citenamefont {Tomadin}, \citenamefont {Manzoni}, \citenamefont {Kim},
  \citenamefont {Lombardo}, \citenamefont {Milana}, \citenamefont {Nair},
  \citenamefont {Novoselov}, \citenamefont {Ferrari}, \citenamefont {Cerullo},\
  and\ \citenamefont {Polini}}]{Natcomm2013brida}%
  \BibitemOpen
  \bibfield  {author} {\bibinfo {author} {\bibfnamefont {D.}~\bibnamefont
  {Brida}}, \bibinfo {author} {\bibfnamefont {A.}~\bibnamefont {Tomadin}},
  \bibinfo {author} {\bibfnamefont {C.}~\bibnamefont {Manzoni}}, \bibinfo
  {author} {\bibfnamefont {Y.~J.}\ \bibnamefont {Kim}}, \bibinfo {author}
  {\bibfnamefont {A.}~\bibnamefont {Lombardo}}, \bibinfo {author}
  {\bibfnamefont {S.}~\bibnamefont {Milana}}, \bibinfo {author} {\bibfnamefont
  {R.~R.}\ \bibnamefont {Nair}}, \bibinfo {author} {\bibfnamefont {K.~S.}\
  \bibnamefont {Novoselov}}, \bibinfo {author} {\bibfnamefont {A.~C.}\
  \bibnamefont {Ferrari}}, \bibinfo {author} {\bibfnamefont {G.}~\bibnamefont
  {Cerullo}}, \ and\ \bibinfo {author} {\bibfnamefont {M.}~\bibnamefont
  {Polini}},\ }\href@noop {} {\bibfield  {journal} {\bibinfo  {journal} {Nat.
  Commun.}\ }\textbf {\bibinfo {volume} {4}},\ \bibinfo {pages} {1987}
  (\bibinfo {year} {2013})}\BibitemShut {NoStop}%
\bibitem [{\citenamefont {Song}\ \emph {et~al.}(2013)\citenamefont {Song},
  \citenamefont {Tielrooij}, \citenamefont {Koppens},\ and\ \citenamefont
  {Levitov}}]{PRB2013song}%
  \BibitemOpen
  \bibfield  {author} {\bibinfo {author} {\bibfnamefont {J.~C.~W.}\
  \bibnamefont {Song}}, \bibinfo {author} {\bibfnamefont {K.~J.}\ \bibnamefont
  {Tielrooij}}, \bibinfo {author} {\bibfnamefont {F.~H.~L.}\ \bibnamefont
  {Koppens}}, \ and\ \bibinfo {author} {\bibfnamefont {L.~S.}\ \bibnamefont
  {Levitov}},\ }\href {\doibase 10.1103/PhysRevB.87.155429} {\bibfield
  {journal} {\bibinfo  {journal} {Phys. Rev. B}\ }\textbf {\bibinfo {volume}
  {87}},\ \bibinfo {pages} {155429} (\bibinfo {year} {2013})}\BibitemShut
  {NoStop}%
\bibitem [{\citenamefont {Mics}\ \emph {et~al.}(2015)\citenamefont {Mics},
  \citenamefont {Tielrooij}, \citenamefont {Parvez}, \citenamefont {Jensen},
  \citenamefont {Ivanov}, \citenamefont {Feng}, \citenamefont {Mullen},
  \citenamefont {Bonn},\ and\ \citenamefont {Turchinovich}}]{Natcomm2015mics}%
  \BibitemOpen
  \bibfield  {author} {\bibinfo {author} {\bibfnamefont {Z.}~\bibnamefont
  {Mics}}, \bibinfo {author} {\bibfnamefont {K.-J.}\ \bibnamefont {Tielrooij}},
  \bibinfo {author} {\bibfnamefont {K.}~\bibnamefont {Parvez}}, \bibinfo
  {author} {\bibfnamefont {S.~A.}\ \bibnamefont {Jensen}}, \bibinfo {author}
  {\bibfnamefont {I.}~\bibnamefont {Ivanov}}, \bibinfo {author} {\bibfnamefont
  {X.}~\bibnamefont {Feng}}, \bibinfo {author} {\bibfnamefont {K.}~\bibnamefont
  {Mullen}}, \bibinfo {author} {\bibfnamefont {M.}~\bibnamefont {Bonn}}, \ and\
  \bibinfo {author} {\bibfnamefont {D.}~\bibnamefont {Turchinovich}},\
  }\href@noop {} {\bibfield  {journal} {\bibinfo  {journal} {Nat Commun}\
  }\textbf {\bibinfo {volume} {6}},\ \bibinfo {pages} {7655} (\bibinfo {year}
  {2015})}\BibitemShut {NoStop}%
\bibitem [{\citenamefont {Mihnev}\ \emph {et~al.}(2016)\citenamefont {Mihnev},
  \citenamefont {Kadi}, \citenamefont {Divin}, \citenamefont {Winzer},
  \citenamefont {Lee}, \citenamefont {Liu}, \citenamefont {Zhong},
  \citenamefont {Berger}, \citenamefont {de~Heer}, \citenamefont {Malic},
  \citenamefont {Knorr},\ and\ \citenamefont {Norris}}]{Natcomm2016mihnev}%
  \BibitemOpen
  \bibfield  {author} {\bibinfo {author} {\bibfnamefont {M.~T.}\ \bibnamefont
  {Mihnev}}, \bibinfo {author} {\bibfnamefont {F.}~\bibnamefont {Kadi}},
  \bibinfo {author} {\bibfnamefont {C.~J.}\ \bibnamefont {Divin}}, \bibinfo
  {author} {\bibfnamefont {T.}~\bibnamefont {Winzer}}, \bibinfo {author}
  {\bibfnamefont {S.}~\bibnamefont {Lee}}, \bibinfo {author} {\bibfnamefont
  {C.-H.}\ \bibnamefont {Liu}}, \bibinfo {author} {\bibfnamefont
  {Z.}~\bibnamefont {Zhong}}, \bibinfo {author} {\bibfnamefont
  {C.}~\bibnamefont {Berger}}, \bibinfo {author} {\bibfnamefont {W.~A.}\
  \bibnamefont {de~Heer}}, \bibinfo {author} {\bibfnamefont {E.}~\bibnamefont
  {Malic}}, \bibinfo {author} {\bibfnamefont {A.}~\bibnamefont {Knorr}}, \ and\
  \bibinfo {author} {\bibfnamefont {T.~B.}\ \bibnamefont {Norris}},\
  }\href@noop {} {\bibfield  {journal} {\bibinfo  {journal} {Nature
  Communications}\ }\textbf {\bibinfo {volume} {7}},\ \bibinfo {pages} {11617}
  (\bibinfo {year} {2016})}\BibitemShut {NoStop}%
\bibitem [{\citenamefont {Malic}\ \emph {et~al.}(2012)\citenamefont {Malic},
  \citenamefont {Winzer},\ and\ \citenamefont {Knorr}}]{APL2012malic}%
  \BibitemOpen
  \bibfield  {author} {\bibinfo {author} {\bibfnamefont {E.}~\bibnamefont
  {Malic}}, \bibinfo {author} {\bibfnamefont {T.}~\bibnamefont {Winzer}}, \
  and\ \bibinfo {author} {\bibfnamefont {A.}~\bibnamefont {Knorr}},\
  }\href@noop {} {\bibfield  {journal} {\bibinfo  {journal} {Applied Physics
  Letters}\ }\textbf {\bibinfo {volume} {101}},\ \bibinfo {eid} {213110}
  (\bibinfo {year} {2012})}\BibitemShut {NoStop}%
\bibitem [{\citenamefont {Mittendorff}\ \emph {et~al.}(2014)\citenamefont
  {Mittendorff}, \citenamefont {Winzer}, \citenamefont {Malic}, \citenamefont
  {Knorr}, \citenamefont {Berger}, \citenamefont {de~Heer}, \citenamefont
  {Schneider}, \citenamefont {Helm},\ and\ \citenamefont
  {Winnerl}}]{Nanolett2014mittendorff}%
  \BibitemOpen
  \bibfield  {author} {\bibinfo {author} {\bibfnamefont {M.}~\bibnamefont
  {Mittendorff}}, \bibinfo {author} {\bibfnamefont {T.}~\bibnamefont {Winzer}},
  \bibinfo {author} {\bibfnamefont {E.}~\bibnamefont {Malic}}, \bibinfo
  {author} {\bibfnamefont {A.}~\bibnamefont {Knorr}}, \bibinfo {author}
  {\bibfnamefont {C.}~\bibnamefont {Berger}}, \bibinfo {author} {\bibfnamefont
  {W.~A.}\ \bibnamefont {de~Heer}}, \bibinfo {author} {\bibfnamefont
  {H.}~\bibnamefont {Schneider}}, \bibinfo {author} {\bibfnamefont
  {M.}~\bibnamefont {Helm}}, \ and\ \bibinfo {author} {\bibfnamefont
  {S.}~\bibnamefont {Winnerl}},\ }\href@noop {} {\bibfield  {journal} {\bibinfo
   {journal} {Nano Letters}\ }\textbf {\bibinfo {volume} {14}},\ \bibinfo
  {pages} {1504} (\bibinfo {year} {2014})}\BibitemShut {NoStop}%
\bibitem [{\citenamefont {Trushin}\ \emph {et~al.}(2015)\citenamefont
  {Trushin}, \citenamefont {Grupp}, \citenamefont {Soavi}, \citenamefont
  {Budweg}, \citenamefont {De~Fazio}, \citenamefont {Sassi}, \citenamefont
  {Lombardo}, \citenamefont {Ferrari}, \citenamefont {Belzig}, \citenamefont
  {Leitenstorfer},\ and\ \citenamefont {Brida}}]{PRB2015trushin}%
  \BibitemOpen
  \bibfield  {author} {\bibinfo {author} {\bibfnamefont {M.}~\bibnamefont
  {Trushin}}, \bibinfo {author} {\bibfnamefont {A.}~\bibnamefont {Grupp}},
  \bibinfo {author} {\bibfnamefont {G.}~\bibnamefont {Soavi}}, \bibinfo
  {author} {\bibfnamefont {A.}~\bibnamefont {Budweg}}, \bibinfo {author}
  {\bibfnamefont {D.}~\bibnamefont {De~Fazio}}, \bibinfo {author}
  {\bibfnamefont {U.}~\bibnamefont {Sassi}}, \bibinfo {author} {\bibfnamefont
  {A.}~\bibnamefont {Lombardo}}, \bibinfo {author} {\bibfnamefont {A.~C.}\
  \bibnamefont {Ferrari}}, \bibinfo {author} {\bibfnamefont {W.}~\bibnamefont
  {Belzig}}, \bibinfo {author} {\bibfnamefont {A.}~\bibnamefont
  {Leitenstorfer}}, \ and\ \bibinfo {author} {\bibfnamefont {D.}~\bibnamefont
  {Brida}},\ }\href@noop {} {\bibfield  {journal} {\bibinfo  {journal} {Phys.
  Rev. B}\ }\textbf {\bibinfo {volume} {92}},\ \bibinfo {pages} {165429}
  (\bibinfo {year} {2015})}\BibitemShut {NoStop}%
\bibitem [{\citenamefont {Fritz}\ \emph {et~al.}(2008)\citenamefont {Fritz},
  \citenamefont {Schmalian}, \citenamefont {M\"uller},\ and\ \citenamefont
  {Sachdev}}]{PRB2008fritz}%
  \BibitemOpen
  \bibfield  {author} {\bibinfo {author} {\bibfnamefont {L.}~\bibnamefont
  {Fritz}}, \bibinfo {author} {\bibfnamefont {J.}~\bibnamefont {Schmalian}},
  \bibinfo {author} {\bibfnamefont {M.}~\bibnamefont {M\"uller}}, \ and\
  \bibinfo {author} {\bibfnamefont {S.}~\bibnamefont {Sachdev}},\ }\href@noop
  {} {\bibfield  {journal} {\bibinfo  {journal} {Phys. Rev. B}\ }\textbf
  {\bibinfo {volume} {78}},\ \bibinfo {pages} {085416} (\bibinfo {year}
  {2008})}\BibitemShut {NoStop}%
\bibitem [{\citenamefont {Malic}\ \emph {et~al.}(2011)\citenamefont {Malic},
  \citenamefont {Winzer}, \citenamefont {Bobkin},\ and\ \citenamefont
  {Knorr}}]{PRB2011malic}%
  \BibitemOpen
  \bibfield  {author} {\bibinfo {author} {\bibfnamefont {E.}~\bibnamefont
  {Malic}}, \bibinfo {author} {\bibfnamefont {T.}~\bibnamefont {Winzer}},
  \bibinfo {author} {\bibfnamefont {E.}~\bibnamefont {Bobkin}}, \ and\ \bibinfo
  {author} {\bibfnamefont {A.}~\bibnamefont {Knorr}},\ }\href@noop {}
  {\bibfield  {journal} {\bibinfo  {journal} {Phys. Rev. B}\ }\textbf {\bibinfo
  {volume} {84}},\ \bibinfo {pages} {205406} (\bibinfo {year}
  {2011})}\BibitemShut {NoStop}%
\bibitem [{\citenamefont {Trushin}\ and\ \citenamefont
  {Schliemann}(2011{\natexlab{a}})}]{EPL2011trushin}%
  \BibitemOpen
  \bibfield  {author} {\bibinfo {author} {\bibfnamefont {M.}~\bibnamefont
  {Trushin}}\ and\ \bibinfo {author} {\bibfnamefont {J.}~\bibnamefont
  {Schliemann}},\ }\href@noop {} {\bibfield  {journal} {\bibinfo  {journal}
  {EPL (Europhysics Letters)}\ }\textbf {\bibinfo {volume} {96}},\ \bibinfo
  {pages} {37006} (\bibinfo {year} {2011}{\natexlab{a}})}\BibitemShut {NoStop}%
\bibitem [{\citenamefont {Trushin}\ and\ \citenamefont
  {Schliemann}(2011{\natexlab{b}})}]{PRL2011trushin}%
  \BibitemOpen
  \bibfield  {author} {\bibinfo {author} {\bibfnamefont {M.}~\bibnamefont
  {Trushin}}\ and\ \bibinfo {author} {\bibfnamefont {J.}~\bibnamefont
  {Schliemann}},\ }\href {\doibase 10.1103/PhysRevLett.107.156801} {\bibfield
  {journal} {\bibinfo  {journal} {Phys. Rev. Lett.}\ }\textbf {\bibinfo
  {volume} {107}},\ \bibinfo {pages} {156801} (\bibinfo {year}
  {2011}{\natexlab{b}})}\BibitemShut {NoStop}%
\bibitem [{\citenamefont {Echtermeyer}\ \emph {et~al.}(2014)\citenamefont
  {Echtermeyer}, \citenamefont {Nene}, \citenamefont {Trushin}, \citenamefont
  {Gorbachev}, \citenamefont {Eiden}, \citenamefont {Milana}, \citenamefont
  {Sun}, \citenamefont {Schliemann}, \citenamefont {Lidorikis}, \citenamefont
  {Novoselov},\ and\ \citenamefont {Ferrari}}]{Nanolett2014echtermeyer}%
  \BibitemOpen
  \bibfield  {author} {\bibinfo {author} {\bibfnamefont {T.~J.}\ \bibnamefont
  {Echtermeyer}}, \bibinfo {author} {\bibfnamefont {P.~S.}\ \bibnamefont
  {Nene}}, \bibinfo {author} {\bibfnamefont {M.}~\bibnamefont {Trushin}},
  \bibinfo {author} {\bibfnamefont {R.~V.}\ \bibnamefont {Gorbachev}}, \bibinfo
  {author} {\bibfnamefont {A.~L.}\ \bibnamefont {Eiden}}, \bibinfo {author}
  {\bibfnamefont {S.}~\bibnamefont {Milana}}, \bibinfo {author} {\bibfnamefont
  {Z.}~\bibnamefont {Sun}}, \bibinfo {author} {\bibfnamefont {J.}~\bibnamefont
  {Schliemann}}, \bibinfo {author} {\bibfnamefont {E.}~\bibnamefont
  {Lidorikis}}, \bibinfo {author} {\bibfnamefont {K.~S.}\ \bibnamefont
  {Novoselov}}, \ and\ \bibinfo {author} {\bibfnamefont {A.~C.}\ \bibnamefont
  {Ferrari}},\ }\href@noop {} {\bibfield  {journal} {\bibinfo  {journal} {Nano
  Letters}\ }\textbf {\bibinfo {volume} {14}},\ \bibinfo {pages} {3733}
  (\bibinfo {year} {2014})}\BibitemShut {NoStop}%
\bibitem [{\citenamefont {Jang}\ \emph {et~al.}(2008)\citenamefont {Jang},
  \citenamefont {Adam}, \citenamefont {Chen}, \citenamefont {Williams},
  \citenamefont {Das~Sarma},\ and\ \citenamefont {Fuhrer}}]{PRL2008jang}%
  \BibitemOpen
  \bibfield  {author} {\bibinfo {author} {\bibfnamefont {C.}~\bibnamefont
  {Jang}}, \bibinfo {author} {\bibfnamefont {S.}~\bibnamefont {Adam}}, \bibinfo
  {author} {\bibfnamefont {J.-H.}\ \bibnamefont {Chen}}, \bibinfo {author}
  {\bibfnamefont {E.~D.}\ \bibnamefont {Williams}}, \bibinfo {author}
  {\bibfnamefont {S.}~\bibnamefont {Das~Sarma}}, \ and\ \bibinfo {author}
  {\bibfnamefont {M.~S.}\ \bibnamefont {Fuhrer}},\ }\href {\doibase
  10.1103/PhysRevLett.101.146805} {\bibfield  {journal} {\bibinfo  {journal}
  {Phys. Rev. Lett.}\ }\textbf {\bibinfo {volume} {101}},\ \bibinfo {pages}
  {146805} (\bibinfo {year} {2008})}\BibitemShut {NoStop}%
\bibitem [{\citenamefont {Geick}\ \emph {et~al.}(1966)\citenamefont {Geick},
  \citenamefont {Perry},\ and\ \citenamefont {Rupprecht}}]{PR1966geick}%
  \BibitemOpen
  \bibfield  {author} {\bibinfo {author} {\bibfnamefont {R.}~\bibnamefont
  {Geick}}, \bibinfo {author} {\bibfnamefont {C.~H.}\ \bibnamefont {Perry}}, \
  and\ \bibinfo {author} {\bibfnamefont {G.}~\bibnamefont {Rupprecht}},\ }\href
  {\doibase 10.1103/PhysRev.146.543} {\bibfield  {journal} {\bibinfo  {journal}
  {Phys. Rev.}\ }\textbf {\bibinfo {volume} {146}},\ \bibinfo {pages} {543}
  (\bibinfo {year} {1966})}\BibitemShut {NoStop}%
\bibitem [{\citenamefont {Patrick}\ and\ \citenamefont
  {Choyke}(1970)}]{PRB1970lyle}%
  \BibitemOpen
  \bibfield  {author} {\bibinfo {author} {\bibfnamefont {L.}~\bibnamefont
  {Patrick}}\ and\ \bibinfo {author} {\bibfnamefont {W.~J.}\ \bibnamefont
  {Choyke}},\ }\href {\doibase 10.1103/PhysRevB.2.2255} {\bibfield  {journal}
  {\bibinfo  {journal} {Phys. Rev. B}\ }\textbf {\bibinfo {volume} {2}},\
  \bibinfo {pages} {2255} (\bibinfo {year} {1970})}\BibitemShut {NoStop}%
\bibitem [{\citenamefont {Dawlaty}\ \emph {et~al.}(2008)\citenamefont
  {Dawlaty}, \citenamefont {Shivaraman}, \citenamefont {Chandrashekhar},
  \citenamefont {Rana},\ and\ \citenamefont {Spencer}}]{APL2008dawlaty}%
  \BibitemOpen
  \bibfield  {author} {\bibinfo {author} {\bibfnamefont {J.~M.}\ \bibnamefont
  {Dawlaty}}, \bibinfo {author} {\bibfnamefont {S.}~\bibnamefont {Shivaraman}},
  \bibinfo {author} {\bibfnamefont {M.}~\bibnamefont {Chandrashekhar}},
  \bibinfo {author} {\bibfnamefont {F.}~\bibnamefont {Rana}}, \ and\ \bibinfo
  {author} {\bibfnamefont {M.~G.}\ \bibnamefont {Spencer}},\ }\href@noop {}
  {\bibfield  {journal} {\bibinfo  {journal} {Applied Physics Letters}\
  }\textbf {\bibinfo {volume} {92}},\ \bibinfo {pages} {042116} (\bibinfo
  {year} {2008})}\BibitemShut {NoStop}%
\bibitem [{\citenamefont {Lui}\ \emph {et~al.}(2010)\citenamefont {Lui},
  \citenamefont {Mak}, \citenamefont {Shan},\ and\ \citenamefont
  {Heinz}}]{PRL2010lui}%
  \BibitemOpen
  \bibfield  {author} {\bibinfo {author} {\bibfnamefont {C.~H.}\ \bibnamefont
  {Lui}}, \bibinfo {author} {\bibfnamefont {K.~F.}\ \bibnamefont {Mak}},
  \bibinfo {author} {\bibfnamefont {J.}~\bibnamefont {Shan}}, \ and\ \bibinfo
  {author} {\bibfnamefont {T.~F.}\ \bibnamefont {Heinz}},\ }\href {\doibase
  10.1103/PhysRevLett.105.127404} {\bibfield  {journal} {\bibinfo  {journal}
  {Phys. Rev. Lett.}\ }\textbf {\bibinfo {volume} {105}},\ \bibinfo {pages}
  {127404} (\bibinfo {year} {2010})}\BibitemShut {NoStop}%
\bibitem [{\citenamefont {Breusing}\ \emph {et~al.}(2011)\citenamefont
  {Breusing}, \citenamefont {Kuehn}, \citenamefont {Winzer}, \citenamefont
  {Mali\ifmmode~\acute{c}\else \'{c}\fi{}}, \citenamefont {Milde},
  \citenamefont {Severin}, \citenamefont {Rabe}, \citenamefont {Ropers},
  \citenamefont {Knorr},\ and\ \citenamefont {Elsaesser}}]{PRB2011breusing}%
  \BibitemOpen
  \bibfield  {author} {\bibinfo {author} {\bibfnamefont {M.}~\bibnamefont
  {Breusing}}, \bibinfo {author} {\bibfnamefont {S.}~\bibnamefont {Kuehn}},
  \bibinfo {author} {\bibfnamefont {T.}~\bibnamefont {Winzer}}, \bibinfo
  {author} {\bibfnamefont {E.}~\bibnamefont {Mali\ifmmode~\acute{c}\else
  \'{c}\fi{}}}, \bibinfo {author} {\bibfnamefont {F.}~\bibnamefont {Milde}},
  \bibinfo {author} {\bibfnamefont {N.}~\bibnamefont {Severin}}, \bibinfo
  {author} {\bibfnamefont {J.~P.}\ \bibnamefont {Rabe}}, \bibinfo {author}
  {\bibfnamefont {C.}~\bibnamefont {Ropers}}, \bibinfo {author} {\bibfnamefont
  {A.}~\bibnamefont {Knorr}}, \ and\ \bibinfo {author} {\bibfnamefont
  {T.}~\bibnamefont {Elsaesser}},\ }\href@noop {} {\bibfield  {journal}
  {\bibinfo  {journal} {Phys. Rev. B}\ }\textbf {\bibinfo {volume} {83}},\
  \bibinfo {pages} {153410} (\bibinfo {year} {2011})}\BibitemShut {NoStop}%
\bibitem [{\citenamefont {Hale}\ \emph {et~al.}(2011)\citenamefont {Hale},
  \citenamefont {Hornett}, \citenamefont {Moger}, \citenamefont {Horsell},\
  and\ \citenamefont {Hendry}}]{PRB2011hale}%
  \BibitemOpen
  \bibfield  {author} {\bibinfo {author} {\bibfnamefont {P.~J.}\ \bibnamefont
  {Hale}}, \bibinfo {author} {\bibfnamefont {S.~M.}\ \bibnamefont {Hornett}},
  \bibinfo {author} {\bibfnamefont {J.}~\bibnamefont {Moger}}, \bibinfo
  {author} {\bibfnamefont {D.~W.}\ \bibnamefont {Horsell}}, \ and\ \bibinfo
  {author} {\bibfnamefont {E.}~\bibnamefont {Hendry}},\ }\href {\doibase
  10.1103/PhysRevB.83.121404} {\bibfield  {journal} {\bibinfo  {journal} {Phys.
  Rev. B}\ }\textbf {\bibinfo {volume} {83}},\ \bibinfo {pages} {121404}
  (\bibinfo {year} {2011})}\BibitemShut {NoStop}%
\bibitem [{\citenamefont {Shang}\ \emph {et~al.}(2011)\citenamefont {Shang},
  \citenamefont {Yu}, \citenamefont {Lin},\ and\ \citenamefont
  {Gurzadyan}}]{ACSnano2011shang}%
  \BibitemOpen
  \bibfield  {author} {\bibinfo {author} {\bibfnamefont {J.}~\bibnamefont
  {Shang}}, \bibinfo {author} {\bibfnamefont {T.}~\bibnamefont {Yu}}, \bibinfo
  {author} {\bibfnamefont {J.}~\bibnamefont {Lin}}, \ and\ \bibinfo {author}
  {\bibfnamefont {G.~G.}\ \bibnamefont {Gurzadyan}},\ }\href@noop {} {\bibfield
   {journal} {\bibinfo  {journal} {ACS Nano}\ }\textbf {\bibinfo {volume}
  {5}},\ \bibinfo {pages} {3278} (\bibinfo {year} {2011})}\BibitemShut
  {NoStop}%
\bibitem [{\citenamefont {Yan}\ \emph {et~al.}(2014)\citenamefont {Yan},
  \citenamefont {Yao}, \citenamefont {Liu}, \citenamefont {Zhao}, \citenamefont
  {Chen}, \citenamefont {Gao}, \citenamefont {Xin}, \citenamefont {Chen},\ and\
  \citenamefont {Tian}}]{PRB2014yan}%
  \BibitemOpen
  \bibfield  {author} {\bibinfo {author} {\bibfnamefont {X.-Q.}\ \bibnamefont
  {Yan}}, \bibinfo {author} {\bibfnamefont {J.}~\bibnamefont {Yao}}, \bibinfo
  {author} {\bibfnamefont {Z.-B.}\ \bibnamefont {Liu}}, \bibinfo {author}
  {\bibfnamefont {X.}~\bibnamefont {Zhao}}, \bibinfo {author} {\bibfnamefont
  {X.-D.}\ \bibnamefont {Chen}}, \bibinfo {author} {\bibfnamefont
  {C.}~\bibnamefont {Gao}}, \bibinfo {author} {\bibfnamefont {W.}~\bibnamefont
  {Xin}}, \bibinfo {author} {\bibfnamefont {Y.}~\bibnamefont {Chen}}, \ and\
  \bibinfo {author} {\bibfnamefont {J.-G.}\ \bibnamefont {Tian}},\ }\href@noop
  {} {\bibfield  {journal} {\bibinfo  {journal} {Phys. Rev. B}\ }\textbf
  {\bibinfo {volume} {90}},\ \bibinfo {pages} {134308} (\bibinfo {year}
  {2014})}\BibitemShut {NoStop}%
\bibitem [{\citenamefont {Gierz}\ \emph {et~al.}(2015)\citenamefont {Gierz},
  \citenamefont {Calegari}, \citenamefont {Aeschlimann}, \citenamefont
  {Ch\'avez~Cervantes}, \citenamefont {Cacho}, \citenamefont {Chapman},
  \citenamefont {Springate}, \citenamefont {Link}, \citenamefont {Starke},
  \citenamefont {Ast},\ and\ \citenamefont {Cavalleri}}]{PRL2015gierz}%
  \BibitemOpen
  \bibfield  {author} {\bibinfo {author} {\bibfnamefont {I.}~\bibnamefont
  {Gierz}}, \bibinfo {author} {\bibfnamefont {F.}~\bibnamefont {Calegari}},
  \bibinfo {author} {\bibfnamefont {S.}~\bibnamefont {Aeschlimann}}, \bibinfo
  {author} {\bibfnamefont {M.}~\bibnamefont {Ch\'avez~Cervantes}}, \bibinfo
  {author} {\bibfnamefont {C.}~\bibnamefont {Cacho}}, \bibinfo {author}
  {\bibfnamefont {R.~T.}\ \bibnamefont {Chapman}}, \bibinfo {author}
  {\bibfnamefont {E.}~\bibnamefont {Springate}}, \bibinfo {author}
  {\bibfnamefont {S.}~\bibnamefont {Link}}, \bibinfo {author} {\bibfnamefont
  {U.}~\bibnamefont {Starke}}, \bibinfo {author} {\bibfnamefont {C.~R.}\
  \bibnamefont {Ast}}, \ and\ \bibinfo {author} {\bibfnamefont
  {A.}~\bibnamefont {Cavalleri}},\ }\href@noop {} {\bibfield  {journal}
  {\bibinfo  {journal} {Phys. Rev. Lett.}\ }\textbf {\bibinfo {volume} {115}},\
  \bibinfo {pages} {086803} (\bibinfo {year} {2015})}\BibitemShut {NoStop}%
\bibitem [{\citenamefont {Bistritzer}\ and\ \citenamefont
  {MacDonald}(2009)}]{PRL2009bistritzer}%
  \BibitemOpen
  \bibfield  {author} {\bibinfo {author} {\bibfnamefont {R.}~\bibnamefont
  {Bistritzer}}\ and\ \bibinfo {author} {\bibfnamefont {A.~H.}\ \bibnamefont
  {MacDonald}},\ }\href@noop {} {\bibfield  {journal} {\bibinfo  {journal}
  {Phys. Rev. Lett.}\ }\textbf {\bibinfo {volume} {102}},\ \bibinfo {pages}
  {206410} (\bibinfo {year} {2009})}\BibitemShut {NoStop}%
\bibitem [{\citenamefont {Goykhman}\ \emph {et~al.}(2016)\citenamefont
  {Goykhman}, \citenamefont {Sassi}, \citenamefont {Desiatov}, \citenamefont
  {Mazurski}, \citenamefont {Milana}, \citenamefont {de~Fazio}, \citenamefont
  {Eiden}, \citenamefont {Khurgin}, \citenamefont {Shappir}, \citenamefont
  {Levy},\ and\ \citenamefont {Ferrari}}]{NanoLett2016goykhman}%
  \BibitemOpen
  \bibfield  {author} {\bibinfo {author} {\bibfnamefont {I.}~\bibnamefont
  {Goykhman}}, \bibinfo {author} {\bibfnamefont {U.}~\bibnamefont {Sassi}},
  \bibinfo {author} {\bibfnamefont {B.}~\bibnamefont {Desiatov}}, \bibinfo
  {author} {\bibfnamefont {N.}~\bibnamefont {Mazurski}}, \bibinfo {author}
  {\bibfnamefont {S.}~\bibnamefont {Milana}}, \bibinfo {author} {\bibfnamefont
  {D.}~\bibnamefont {de~Fazio}}, \bibinfo {author} {\bibfnamefont
  {A.}~\bibnamefont {Eiden}}, \bibinfo {author} {\bibfnamefont
  {J.}~\bibnamefont {Khurgin}}, \bibinfo {author} {\bibfnamefont
  {J.}~\bibnamefont {Shappir}}, \bibinfo {author} {\bibfnamefont
  {U.}~\bibnamefont {Levy}}, \ and\ \bibinfo {author} {\bibfnamefont {A.~C.}\
  \bibnamefont {Ferrari}},\ }\href@noop {} {\bibfield  {journal} {\bibinfo
  {journal} {Nano Letters}\ }\textbf {\bibinfo {volume} {16}},\ \bibinfo
  {pages} {3005} (\bibinfo {year} {2016})}\BibitemShut {NoStop}%
\bibitem [{\citenamefont {Ma}\ \emph {et~al.}(2016)\citenamefont {Ma},
  \citenamefont {Andersen}, \citenamefont {Nair}, \citenamefont {Gabor},
  \citenamefont {Massicotte}, \citenamefont {Lui}, \citenamefont {Young},
  \citenamefont {Fang}, \citenamefont {Watanabe}, \citenamefont {Taniguchi},
  \citenamefont {Kong}, \citenamefont {Gedik}, \citenamefont {Koppens},\ and\
  \citenamefont {Jarillo-Herrero}}]{NatPhys2016ma}%
  \BibitemOpen
  \bibfield  {author} {\bibinfo {author} {\bibfnamefont {Q.}~\bibnamefont
  {Ma}}, \bibinfo {author} {\bibfnamefont {T.~I.}\ \bibnamefont {Andersen}},
  \bibinfo {author} {\bibfnamefont {N.~L.}\ \bibnamefont {Nair}}, \bibinfo
  {author} {\bibfnamefont {N.~M.}\ \bibnamefont {Gabor}}, \bibinfo {author}
  {\bibfnamefont {M.}~\bibnamefont {Massicotte}}, \bibinfo {author}
  {\bibfnamefont {C.~H.}\ \bibnamefont {Lui}}, \bibinfo {author} {\bibfnamefont
  {A.~F.}\ \bibnamefont {Young}}, \bibinfo {author} {\bibfnamefont
  {W.}~\bibnamefont {Fang}}, \bibinfo {author} {\bibfnamefont {K.}~\bibnamefont
  {Watanabe}}, \bibinfo {author} {\bibfnamefont {T.}~\bibnamefont {Taniguchi}},
  \bibinfo {author} {\bibfnamefont {J.}~\bibnamefont {Kong}}, \bibinfo {author}
  {\bibfnamefont {N.}~\bibnamefont {Gedik}}, \bibinfo {author} {\bibfnamefont
  {F.~H.~L.}\ \bibnamefont {Koppens}}, \ and\ \bibinfo {author} {\bibfnamefont
  {P.}~\bibnamefont {Jarillo-Herrero}},\ }\href@noop {} {\bibfield  {journal}
  {\bibinfo  {journal} {Nat Phys}\ }\textbf {\bibinfo {volume} {12}},\ \bibinfo
  {pages} {455} (\bibinfo {year} {2016})}\BibitemShut {NoStop}%
\bibitem [{\citenamefont {Rodriguez-Nieva}\ \emph {et~al.}(2015)\citenamefont
  {Rodriguez-Nieva}, \citenamefont {Dresselhaus},\ and\ \citenamefont
  {Levitov}}]{NanoLett2015rodriguez}%
  \BibitemOpen
  \bibfield  {author} {\bibinfo {author} {\bibfnamefont {J.~F.}\ \bibnamefont
  {Rodriguez-Nieva}}, \bibinfo {author} {\bibfnamefont {M.~S.}\ \bibnamefont
  {Dresselhaus}}, \ and\ \bibinfo {author} {\bibfnamefont {L.~S.}\ \bibnamefont
  {Levitov}},\ }\href@noop {} {\bibfield  {journal} {\bibinfo  {journal} {Nano
  Letters}\ }\textbf {\bibinfo {volume} {15}},\ \bibinfo {pages} {1451}
  (\bibinfo {year} {2015})}\BibitemShut {NoStop}%
\bibitem [{\citenamefont {Bonaccorso}\ \emph {et~al.}(2010)\citenamefont
  {Bonaccorso}, \citenamefont {Sun}, \citenamefont {Hasan},\ and\ \citenamefont
  {Ferrari}}]{Bonaccorso2010review}%
  \BibitemOpen
  \bibfield  {author} {\bibinfo {author} {\bibfnamefont {F.}~\bibnamefont
  {Bonaccorso}}, \bibinfo {author} {\bibfnamefont {Z.}~\bibnamefont {Sun}},
  \bibinfo {author} {\bibfnamefont {T.}~\bibnamefont {Hasan}}, \ and\ \bibinfo
  {author} {\bibfnamefont {A.~C.}\ \bibnamefont {Ferrari}},\ }\href@noop {}
  {\bibfield  {journal} {\bibinfo  {journal} {Nat. Photon.}\ }\textbf {\bibinfo
  {volume} {4}},\ \bibinfo {pages} {611} (\bibinfo {year} {2010})}\BibitemShut
  {NoStop}%
\bibitem [{\citenamefont {Koppens}\ \emph {et~al.}(2014)\citenamefont
  {Koppens}, \citenamefont {Mueller}, \citenamefont {Avouris}, \citenamefont
  {Ferrari}, \citenamefont {Vitiello},\ and\ \citenamefont
  {Polini}}]{Natnano2014review}%
  \BibitemOpen
  \bibfield  {author} {\bibinfo {author} {\bibfnamefont {F.~H.~L.}\
  \bibnamefont {Koppens}}, \bibinfo {author} {\bibfnamefont {T.}~\bibnamefont
  {Mueller}}, \bibinfo {author} {\bibfnamefont {P.}~\bibnamefont {Avouris}},
  \bibinfo {author} {\bibfnamefont {A.~C.}\ \bibnamefont {Ferrari}}, \bibinfo
  {author} {\bibfnamefont {M.~S.}\ \bibnamefont {Vitiello}}, \ and\ \bibinfo
  {author} {\bibfnamefont {M.}~\bibnamefont {Polini}},\ }\href@noop {}
  {\bibfield  {journal} {\bibinfo  {journal} {Nat Nano}\ }\textbf {\bibinfo
  {volume} {9}},\ \bibinfo {pages} {780} (\bibinfo {year} {2014})}\BibitemShut
  {NoStop}%
\end{thebibliography}%

\end{document}